\documentclass[twocolumn,pra,aps]{revtex4}

\usepackage{mathptmx}
\usepackage{subfigure}
\usepackage{dcolumn}
\usepackage{amsmath,amssymb}
\usepackage{bm}
\usepackage{color}
\usepackage{latexsym}
\usepackage{epstopdf}
\usepackage{color}
\usepackage[english]{babel}
\usepackage{latexsym}
\usepackage{stmaryrd}

\usepackage{psfrag,graphicx} 
\usepackage{epsf} 
\usepackage{subfigure} 
\usepackage{amsmath} 
\usepackage{amssymb} 
\usepackage{amsfonts}
\usepackage{bm}
\usepackage{natbib}
\usepackage{epstopdf}
\usepackage{appendix}

\definecolor{mygrey}{gray}{0.}
\definecolor{myblue}{rgb}{0.2,0.2,0.8}
\definecolor{myzard}{cmyk}{0,0,0.05,0}
\definecolor{mywhite}{rgb}{1,1,1}
\definecolor{myred}{rgb}{1,0.,0.3}

\usepackage[colorlinks=true,citecolor=myblue,linkcolor=myred]{hyperref}
\def\be{\begin{equation}}
\def\ee{\end{equation}}
\def\ba{\begin{align}}
\def\enda{\end{align}}
\def\bi{\begin{itemize}}
\def\ei{\end{itemize}}

 \def\ee{\mathord{\rm e}}

 \def\ee{\mathord{\rm e}}

\def \be{\begin{equation}}
\def \ee{\end{equation}}
\def \ba{\begin{array}}
\def \ea{\end{array}}
\def \bea{\begin{eqnarray}}
\def \eea{\end{eqnarray}}

\newcommand{\ket}[1]{\left\vert #1 \right\rangle}
\newcommand{\bra}[1]{\left\langle #1 \right\vert}
\newcommand{\ketbra}[2]{\ket{ #1}\bra{ #2} }
\newcommand{\bla}[1]{\left( #1 \right)}
\newcommand{\blb}[1]{\left[ #1 \right]}
\newcommand{\ele}[2]{^{#1} \mbox{#2}}

\begin{document}

\title{Diamond based single molecule magnetic resonance spectroscopy}

\author{Jianming Cai,$^{1,3}$ Fedor Jelezko,$^{2,3}$ Martin B. Plenio,$^{1,3}$ Alex Retzker$^{1,4}$}
\affiliation{$^1$ Institut f\"{u}r Theoretische Physik, Albert-Einstein Allee 11, Universit\"{a}t Ulm, 89069 Ulm, Germany}
\affiliation{$^2$ Institut f\"{u}r Quantenoptik, Albert-Einstein Allee 11, Universit\"{a}t Ulm, 89069 Ulm, Germany}
\affiliation{$^3$ Center for Integrated Quantum Science and Technology, Universit\"{a}t Ulm, 89069 Ulm, Germany}
\affiliation{$^4$ Racah Institute of Physics, The Hebrew University of Jerusalem, Jerusalem 91904, Givat Ram, Israel}

\begin{abstract}
The detection of a nuclear spin in an individual molecule represents a key challenge in physics and biology whose solution has been pursued for many years. The small magnetic moment of a single nucleus and the unavoidable environmental noise present the key obstacles for its realization. Here, we demonstrate theoretically that a single nitrogen-vacancy (NV) center in diamond can be used to construct a nano-scale single molecule spectrometer that is capable of detecting the position and spin state of a single nucleus and can determine the distance and alignment of a nuclear or electron spin pair. The proposed device will find applications in single molecule spectroscopy in chemistry and biology, such as in determining protein structure or monitoring macromolecular motions and can thus provide a tool to help unravelling the microscopic mechanisms underlying bio-molecular function.
\end{abstract}

\date{\today}

\maketitle

\section{Introduction} 

As single nuclei have exceedingly small magnetic moments, a large ensemble (typically $10^{18}$ nuclear spins) is necessary to obtain an observable signal employing methods such as magnetic resonance spectroscopy (NMR). As a consequence, chemical and biological processes have usually been tracked with ensemble measurements, which only provide ensemble averages and distribution information. Single-molecule studies can instead allow one to learn structural information and time trajectories of individual molecules free of natural disorder \cite{Orrit99,Silbey04,Lemke08,Lord10}. The detection of a single nucleus in a single molecule can thus provide various new possibilities for single molecule spectroscopy. Furthermore, single nuclear spins have long coherence time due to their weak coupling with the environment which makes them promising candidates for a qubit or a quantum register. For example, it has been proposed to engineer nitrogen nuclear spin in the molecule of $\ele{14}{N}@\mbox{C}_{60}$ \cite{Twa02,Ben06} or phosphorus donor \cite{Kane98} as a qubit candidate. For these purposes efficient readout of the nitrogen nuclear spin state of a single molecule is crucial.

Nitrogen-vacancy (NV) centers in diamond have been used to construct ultrasensitive nano-scale magnetometers \cite{Maze08,Bal08,Hanson11,Sch11,Hall10,Hall09}, and have led to interesting applications in nano imaging \cite{Cole09,Yacoby11} and biology as well \cite{Mcg11,Chao07,Fu07}. NV centers in diamond benefit from long coherence times at room temperature and highly developed techniques for coherent control, noise decoupling schemes and optical readout of their electron spins. A single NV center can be used to detect single external electron spin \cite{Grotz11} and a proximal nuclear spin (within a few atom shells) \cite{Childress06,Jiang09,Neu10}. It has also been shown that it is possible to detect a strongly coupled nuclear spin pair \cite{Liu10} and measure noise spectra \cite{Byl11,Suter11b,Sar08,Yuge11,Gill12} by applying dynamical decoupling pulses. The schemes based on decoupling pulses may nevertheless suffer various constraints such as the limited achievable repetition rate of pulses and the power requirement. As a consequence the detection of individual nuclear spins in the presence of realistic environments still remains a challenging task. The key obstacles originate firstly from the requirement that coherence times are sufficiently long in order to observe the effect of single nuclei on the NV spin. The second key obstacle lies in the difficulty of distinguishing a distant nuclear spin from other environmental nuclei that couple to the NV spin.

In this work, we address both challenges by continuously driving a single NV electron spin and use it as a probe to measure a specific transition frequency of the target system. The role of continuous driving is two-fold: firstly, it decouples the NV spin from the unwanted influence of a spin bath (which is particularly strong for NV centers located close to the surface \cite{Grotz11}) to achieve sufficiently long coherence times \cite{Tim08,Ber11,Cai11,TimNature,Fan07pra,Fone04,Pas04pra}; secondly, by changing the Rabi frequency of the external driving field we can tune the NV spin to match the target frequency and thus selectively enhance the sensitivity for this specific frequency which allows us to single out the target nucleus. With this scheme, we are able to determine the position of a single nucleus in the presence of a realistic environment. We demonstrate how to use this mechanism to implement quantum non-demolition(QND) measurement of a single nitrogen nuclear spin state in a cage molecule of fullerene ($\ele{14}{N}@\mbox{C}_{60}$). This is achieved by exploiting the fact that the flip-flop process between the NV spin and the target system will either be allowed or prohibited dependent on the nuclear spin state for a specific NV spin initial state. Our proposed detector can be applied straightforwardly to the determination of the distance and alignment of a spin pair. We show that the present model can find applications in single molecule spectroscopy with organic spin labels. It can also be used to witness the creation and recombination of charge separate state in radical pair reactions \cite{Ste89}. We expect that the present diamond-based single molecule probe will find more potential applications in determining protein structure and monitoring chemical (biological) processes, see the examples of molecular motors, ATPase and RNA folding in \cite{Noji97,ZXW07,Nish04}.

\begin{figure}[t]
\begin{center}
\includegraphics[width=8cm]{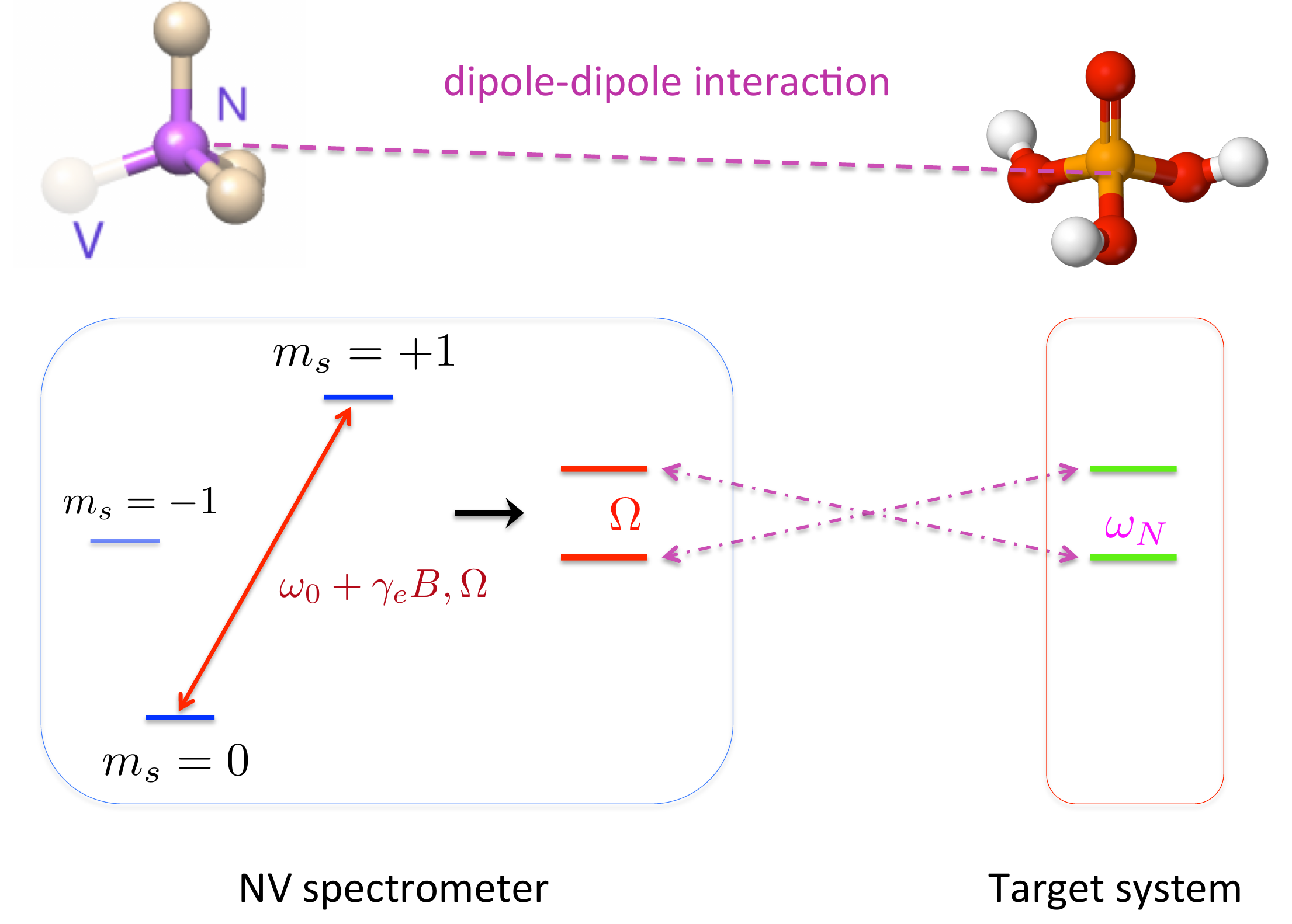}
\end{center}
\caption{(Color online) Basic model of a NV center spectrometer. We apply a continuous field to drive the electronic transition $\ket{m_s=0}\leftrightarrow \ket{m_s=+1}$, which provides an effective dressed spin-$\frac{1}{2}$. When the Rabi frequency $\Omega$ is on resonance with one specific transition frequency of the target system, the flip-flop process will happen between the dressed NV spin and the target system, which leads to a change of the NV dressed state population that can be measured via spin-dependent fluorescence of NV spin. }\label{MODEL}
\end{figure}

\section{Basic model of NV center spectrometer} 
\label{section:basic_model}

The probe in our model is a nitrogen-vacancy (NV) center spin in diamond,  whose ground state is a spin-$1$ with a zero-field splitting of $2.87 \mbox{GHz}$. By applying an additional magnetic field, one can lift the degeneracy of $\ket{m_s=-1}$ and $\ket{m_s=+1}$ and thus allow selective driving with continuous microwave field of one specific electronic transition, e.g. $\ket{m_s=0} \leftrightarrow \ket{m_s=+1}$  (henceforth denoted by $\ket{\downarrow}$ and $\ket{\uparrow}$ respectively). Within the $\ket{\downarrow}$, $\ket{\uparrow}$ subspace the Hamiltonian of the driving field can be written as $H_{NV}= \Omega  \sigma_x $, where $\sigma_x=\ketbra{\uparrow}{\downarrow}+\ketbra{\downarrow}{\uparrow}$ is spin-$\frac{1}{2}$ operator and the eigenstates ($\ket{\uparrow}_x$ and $\ket{\downarrow}_x$) represent the dressed states of the system. We remark that different driving schemes may be used depending on the properties of target systems as we will discuss later, see section \ref{spinstate:sec}. Our goal is to use such a NV dressed spin as a probe to detect a specific frequency in the target system. The magnetic dipole-dipole interaction between the NV spin and another spin is $H_{NV-S}=   \sum_N g_N\blb{3\bla{{\bf S}\cdot \hat{r}_N}\bla{{\bf I}_N\cdot \hat{r}_N}-{\bf S} \cdot {\bf I}_N}$, where ${\bf S}$, ${\bf I}_N$ are the NV spin and target spin operators, and the interaction strength is $g_N=-(\hbar\mu_0 \gamma_N\gamma_e)/(4\pi r_{N}^3)$, wth $\gamma_e$ and $\gamma_N$ the gyromagnetic ratio of electron spin and target spin respectively. The vector $\vec{r}_N=r_N \hat{r}_N$, with the unit vector $\hat{r}_N=(r_N^x,r_N^y,r_N^z)$, connects NV center and the target spin. The large zero-field splitting leads to an energy mismatch which prohibits direct NV spin flip-flop dynamics and allows for the secular approximation to simplify the NV-target spin interaction as follows
\begin{equation}
H_{NV-S} = {\bf S}_z  \sum_N g_N \blb{ 3r_z \bla{r_N^x {\bf I}_N^x + r_N^y {\bf I}_N^y}+\bla{3 (r_N^z)^2-1}{\bf I}_N^z}.
\end{equation}
The coupling operator ${\bf S}_z $ leads to dephasing type interaction in the original NV spin basis but induces flips of NV dressed spin (see Fig.\ref{MODEL}). The flip-flop process will be most efficient when the Rabi frequency $\Omega$, i.e. the energy splitting of the dressed states (($\ket{\uparrow}_x$ or $\ket{\downarrow}_x$)), matches the the transition frequency of the system spin that we wish to probe \cite{Hahn62}. Therefore, one can initialize the NV spin in one of the dressed states and tune the Rabi frequency to determine the transition frequency of the target system by measuring the probability that the NV spin remains in the initial state. The position of the target spin (i.e. the information on the vector $\vec{r}_N$) can be inferred from the flip-flop rate of the NV center if the magnetic moment is known. We remark that continuous driving on one hand achieves the selectively coupling between NV center and the target nuclear spin, on the other hand it decouples NV center from the other species of spins due to the mismatch of the Hartmann-Hahn resonant condition. The residual effect of the environmental spins is suppressed to be on the order of $\sim \delta^2/(\Omega-\omega)$, where $\delta$ represents the magnetic noise from the environmental spin and $\omega$ denotes its Larmor frequency (see the appendix for a simple example which demonstrates the decoupling efficiency of continuous driving).

\begin{figure*}
\begin{center}
\hspace{-0.5cm}
\includegraphics[width=6cm]{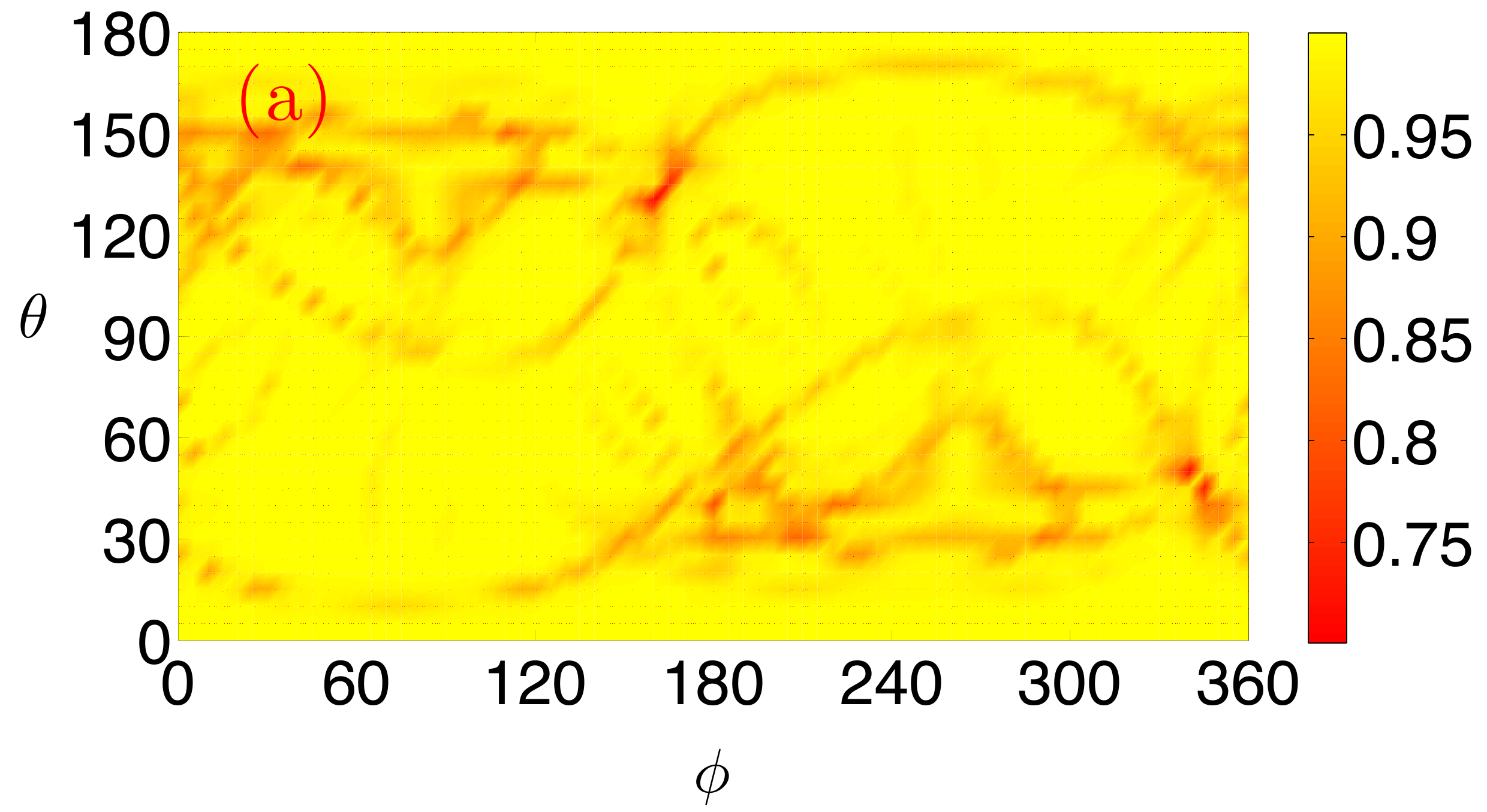}
\includegraphics[width=6cm]{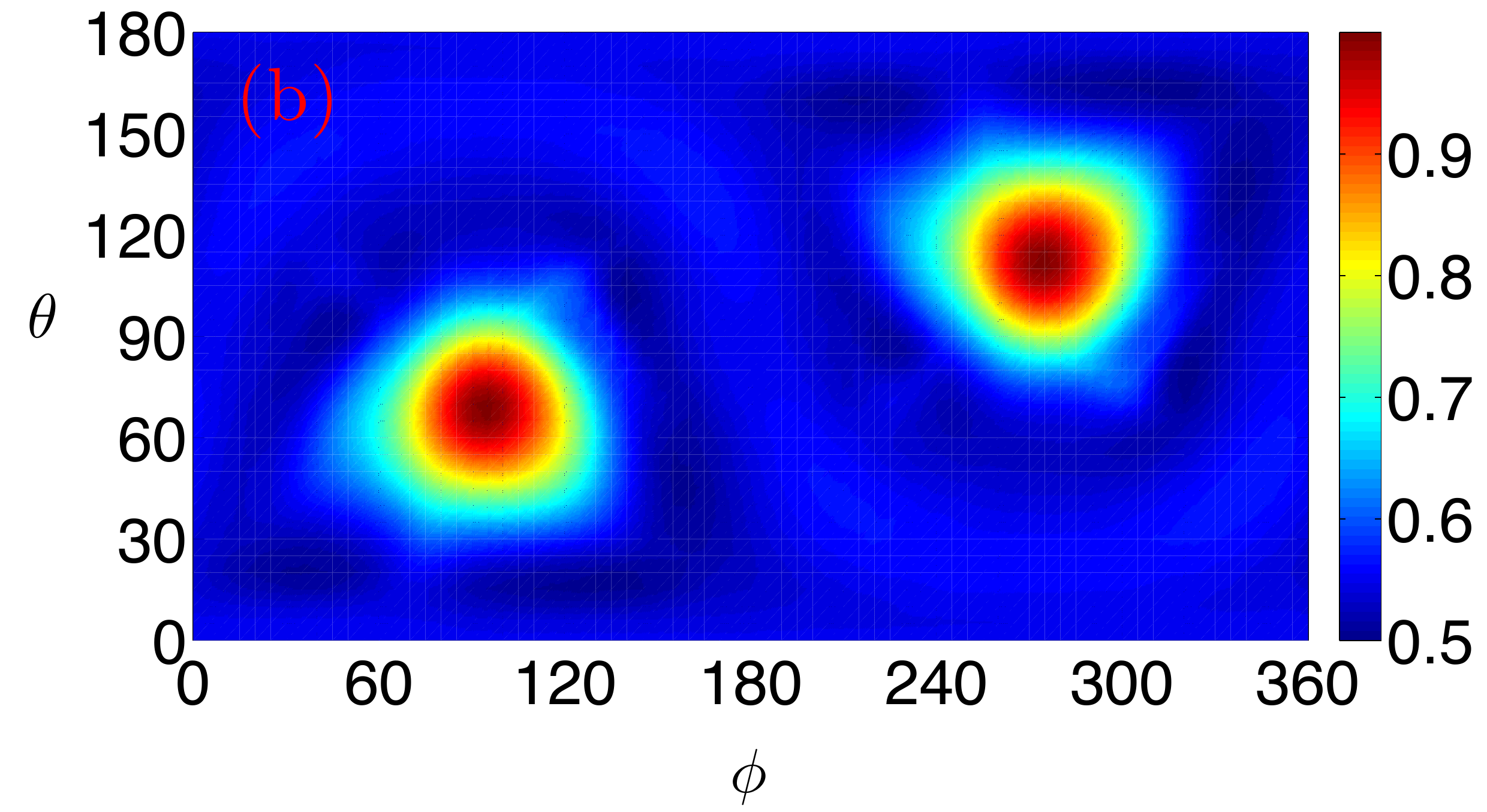}
\includegraphics[width=6cm]{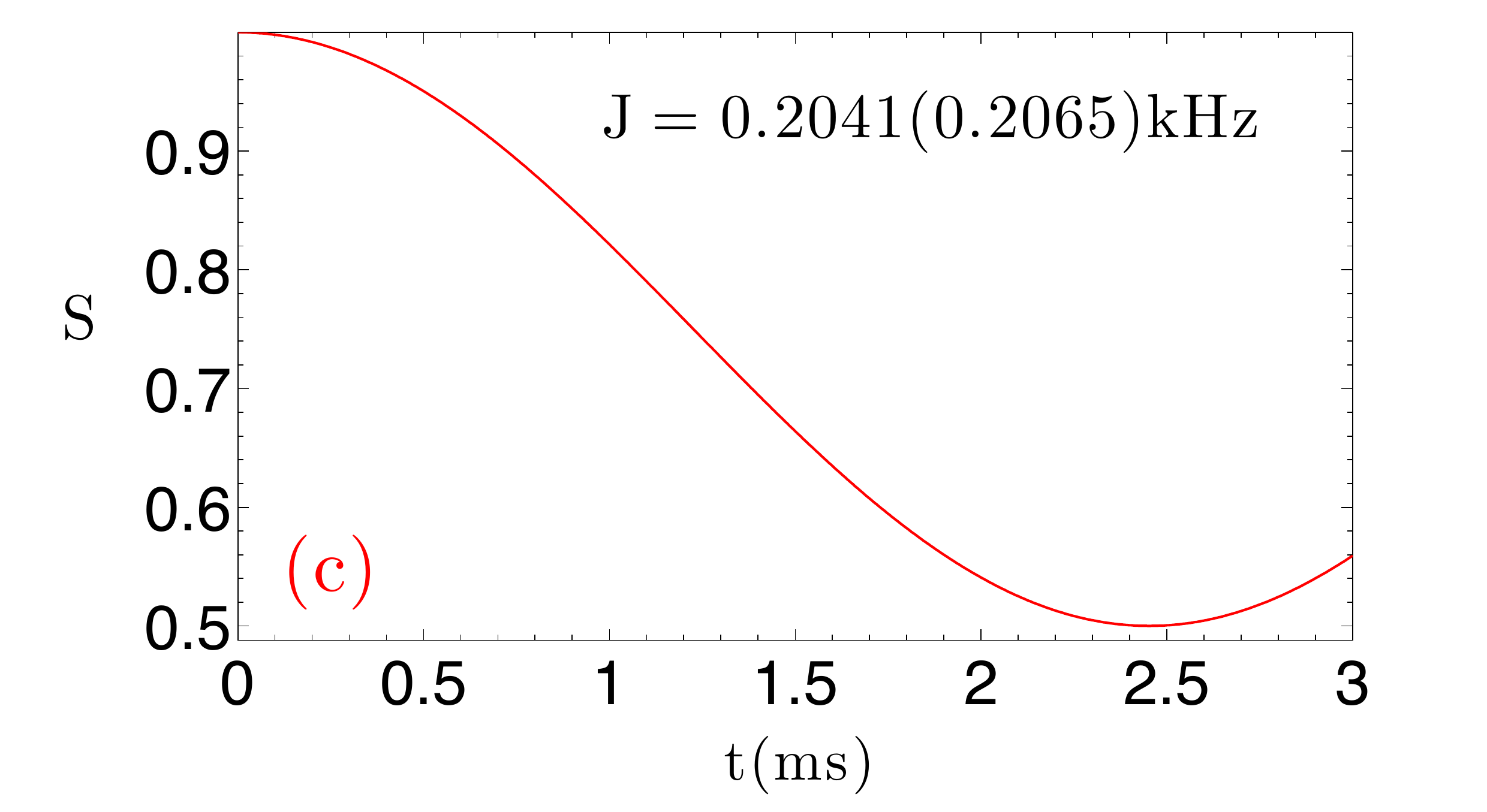}
\end{center}
\caption{(Color online) Measure the position of a single nucleus $\ele{31}{P}$ in a $\ele{1}{H}_3$$\ele{31}{P}$$\mbox{O}_4$ molecule at a distance of $d=5\mbox{nm}$ from the NV center. (a) The signal $S(t)$ measured at time $t=3\mbox{ms}$ for different magnetic field direction as described by $(\theta,\phi)$ without radio frequency driving on resonance with the Larmor frequency of $\ele{1}{H}$. The driving amplitude on NV spin is $\Omega=500 \mbox{kHz}$. (b) The signal $S(t)$ measured at time $t=3\mbox{ms}$ for different magnetic field directions $(\theta,\phi)$ with the application of radio frequency driving ($20 \mbox{kHz}$) on resonance with the Larmor frequency of $\ele{1}{H}$. The driving amplitude on NV spin is tuned to be on resonance with the Larmor frequency of $\ele{31}{P}$ at $\omega=\gamma_{\ele{31}{P}}B=500\mbox{kHz}$ with $B=290\mbox{G}$. The hyperfine vector is in the direction of $\theta_0=68.233^o$ and $\phi_0= 93.841^o$. (b) The signal $S$ as a function of time $t$ with the magnetic field direction chosen to be orthogonal to the hyperfine vector. The estimated value of $J$ is $0.2041 \mbox{kHz}$, which is in good agreement with the exact value $0.2065 \mbox{kHz}$.}\label{EP31}
\end{figure*}

\section{Measure position of a single nucleus} 
\label{pos:sec}

\subsection{Work principles}

We first apply our idea to determine the position of a single nucleus via the flip-flop process between NV dressed spin and the target spin. This is particularly interesting in biology, for example in determining where certain important nuclei (e.g.$\ele{14}{N}$, $\ele{31}{P}$) are located in the protein complex. We introduce the hyperfine vector $\hat{h} (\theta_0,\phi_0)$, which is determined by the unit vector $\hat{r}(r_x,r_y,r_z)$ that connects NV center and the target spin as  
\begin{eqnarray}
\cos{\theta_0}&=& \frac{3 r_z^2-1}{\sqrt{3 r_z^2+1}},\label{ctheta_x}\\
\sin{\theta_0}\cos{\phi_0}&=& \frac{3 r_x r_z}{\sqrt{3 r_z^2+1}},\label{ctheta_y}\\
\sin{\theta_0}\sin{\phi_0}&=& \frac{3 r_y r_z}{\sqrt{3 r_z^2+1}}.\label{ctheta_z}
\end{eqnarray}
The effective total Hamiltonian is thus rewritten as
\begin{equation}
H_E=\Omega {\bf S}_x -\bla{\gamma_N B} {\bf I}_N \cdot  \hat{b}\bla{\theta,\phi} +  \bla{ g_N \sqrt{3 r_z^2+1}} {\bf S}_z  \blb{ \hat{h}\bla{\theta_0,\phi_0}\cdot {\bf I}_N}.
\end{equation}
where $\hat{b}\bla{\theta,\phi}$ denotes the direction of the magnetic field. We note that ${\bf S}_z=\ketbra{+1}{+1}=\frac{1}{2}{I+\sigma_z}$ where $\sigma_z$ is the spin-$\frac{1}{2}$ operator, thus the above Hamiltonian is
\begin{equation}
H_E=\Omega {\bf S}_x -\gamma_N {\bf B}_e\bla{\theta_e,\phi_e} \cdot {\bf I}_N  +  \frac{1}{2}\bla{ g_N \sqrt{3 r_z^2+1}} \sigma_z  \blb{ \hat{h}\bla{\theta_0,\phi_0} \cdot {\bf I}_N},
\end{equation}
where the effective nuclear spin Larmor frequency is given by ${\bf B}_e = {\bf B}(\theta,\phi)-\frac{1}{2}\bla{ g_N \sqrt{3 r_z^2+1}}  \hat{h}\bla{\theta_0,\phi_0}$. We choose the driving amplitude $\Omega_0$ on resonance with the Larmor frequency of the target nuclear spin to satisfy the Hartmann-Hahn matching condition $\Omega = \gamma_N B_e$ \cite{Hahn62}. To measure the position of the target spin, we first prepare NV spin in the initial state $\ket{\uparrow}_x=\frac{1}{\sqrt{2}} \bla{\ket{0}+\ket{1}}$ and assume that the nuclear spin is in the thermal state which is well approximated by $\rho_N(0)=\frac{1}{2} I$ at room temperature. If the effective nuclear spin Larmor frequency is much larger than the NV-nuclear spin coupling (i.e. $\omega_N =\Omega\gg g_N$), the energy unconserved interactions between the NV spin and the nuclear spin is greatly suppressed, and the flip-flop process is dominant. After time $t$, we measure the probability that the NV spin remains in the state $\ket{\uparrow}_x$ which depends on the coupling strength $J$ as follows
\begin{equation}
S(t)=\frac{1}{2}+\frac{1+\cos{(J t)}}{4}.
\label{SIG}
\end{equation}
From the signal $S(t)$ one can infer the value of $J$ that depends on the distance and direction of the nuclear spin to NV center
\begin{equation}
J=\frac{1}{4}\bla{ g_N \sqrt{3 r_z^2+1}} \blb{1-\vert \hat{h}(\theta_0,\phi_0)\cdot \hat{b}(\theta_e,\phi_e)\vert ^2}^{1/2}.
\label{EFJ}
\end{equation}
The spin flip-flop rate $J$ as in Eq.(\ref{EFJ}) depends on the direction of the effective magnetic field $(\theta_e,\phi_e)$. In the case that the applied magnetic field is much larger than the nuclear energy shift induced by NV spin, the direction $(\theta_e,\phi_e)$ can be approximated well by the applied magnetic field direction $\hat{b}(\theta,\phi)$. It can be seen that if and only if the magnetic field direction $\bla{\theta,\phi}$ is in parallel (or anti-parallel) with the hyperfine vector, namely $\theta=\theta_0 (\pi-\theta_0)$ and $\phi=\phi_0 (2\pi-\phi_0)$, the effective flip-flop rate vanishes $J=0$ even though the Hartmann-Hahn resonance condition is still satisfied. If the magnetic field is orthogonal to the hyperfine vector, the flip-flop rate is maximal as $J_m=\frac{1}{4}\bla{ g_N \sqrt{3 r_z^2+1}}$. By choosing a set of magnetic field directions $\hat{b}(\theta,\phi)$, one can obtain the projection of hyperfine interaction along different directions, from which one can derive the direction vector of the target spin ($r_x,r_y,r_z$) with respect to the NV center. We remark that when the applied is not along the NV axis, the eigenstates of NV are in general not the eigenstates of the spin operator $\mathbf{S}_z$. This may affect e.g. the efficiency of NV spin measurement and limit the highest magnetic field applied. One possible way to compensate is to apply additional pulses to map the eigenstates of NV to the one of the spin operator $\mathbf{S}_z$.

We remarked that the fluctuation in the driving field ($\Omega$) will limit the measurement precision. One way to overcome this problem is to use the high-order dressed spin from concatenated dynamical decoupling as proposed in \cite{Cai11} as a probe. For example, we can use a second-order sequential driving scheme with the first-order driving field $H_{d_1}=\Omega_1 \cos(\omega t) \sigma_x$ and a second-order driving field as $H_{d_2}=2 \Omega_2 \cos\bla{\omega t+\frac{\pi}{2}} \cos\bla{\Omega_1 t}\sigma_x$, which can suppress the effect of the first driving field fluctuation \cite{Cai11}. In the second order interaction picture, we can tune the Rabi frequencies to satisfy the Hartmann-Hahn resonant condition as $\Omega_1+\Omega_2 = \omega_N$, and the coupling operator of NV spin with the target spin becomes
\begin{equation}
\sigma_z \rightarrow \frac{1}{2}  \exp{\blb{ i (\Omega_1+\Omega_2)}} \ketbra{+_y} {-_y} +\mbox{h.c.}.
\end{equation}
The energy gap of the second-order dressed spin $\{\ket{+_y},\ket{-_y}\}$ is more robust against the driving field fluctuation \cite{Cai11}, and thus enables our scheme to work with a higher accuracy. With higher order coupling, it is even possible achieve $T_1$ limit coherence time. The required measurement time $t$ is constrained by $T_1$, which can well exceed a few $\mbox{ms}$ for NV centers \cite{Taka08}.

\subsection{Measure a single $\ele{31}{P}$ position in the molecule of $\ele{1}{H}_3 \ele{31}{P} \mbox{O}_4$} 

We demonstrate our idea by showing how to use a NV spin to measure the position of $\ele{31}{P}$ in a single molecule $\ele{1}{H}_3$$\ele{31}{P}$$\mbox{O}_4$. Due to its half-integer nuclear spin and high abundance of $\ele{31}{P}$, NMR spectroscopy based on phosphorus-31 has become a useful tool in studies of biological samples. Phosphorus is commonly found in organic compounds, coordination complexes and proteins, such as phosphatidyl choline which is the major component of lecithin. In $\ele{1}{H}_3$$\ele{31}{P}$$\mbox{O}_4$, $\ele{31}{P}$ interacts with a few adjacent hydrogen atoms, which captures generic features of the protein. Thus it can serve as a paradigmatic example for practical applications of our model. For $\ele{31}{P} $ and $\ele{1}{H}$, the coupling strength is $g=48.6 (6.075) \mbox{kHz}$ for the distance $R_m=0.1 (0.2) \mbox{nm}$, which is relatively strong as compared to the coupling  between NV spin and the target $\ele{31}{P}$ spin. The sole information about the position of $\ele{31}{P}$ is blurred by such a strong interaction, see Fig.\ref{EP31} (a). We can apply a strong magnetic field such that $(\gamma_{\ele{1}{H}}-\gamma_{\ele{31}{P}} )B \gg g_m$, which suppresses the hopping interaction between the target nuclear spins and its neighbouring hydrogen nuclei. The effective Hamiltonian of the target system thus becomes 
\begin{equation}
H_{S}= -\gamma_N {\bf B} \cdot \mathbf{I}_N  - \gamma_m {\bf B} \cdot\sum_m {\bf I}_m +\sum_{m}g_m \bla{1-3 r_z^2}   {\bf I}_m^z {\bf I}_N^z
\label{HamilH3PO4}
\end{equation}
where $\mathbf{I}_N$ and $\mathbf{I}_m$ denote the spin operator of $\ele{31}{P}$ and $\ele{1}{H}$ respectively. To eliminate the line broadening (i.e. the last term Eq.\ref{HamilH3PO4}), we propose to drive the hydrogen nuclei continuously. Since the Larmor frequencies can be made sufficiently different for $\ele{31}{P}$ and $ \ele{1}{H}$, it is possible for us to selectively drive the hydrogen nuclear spins while not affecting the target $\ele{31}{P}$ nucleus due to a large detuning. This leads to the system Hamiltonian as follows
\begin{equation}
H_{S}= -\gamma_N {\bf B} \cdot \mathbf{I}_N + \Omega_1 \cdot\sum_m {\bf I}_m^x + \sum_{m}g_m \bla{1-3 r_z^2}   {\bf I}_m^z {\bf I}_N^z.
\end{equation}
If the Rabi frequency $\Omega_1$ is large enough, the interaction term ${\bf I}_m^z {\bf I}_N^z$ can be effectively averaged out and eliminated. We remark that one can also drive the target $\ele{31}{P}$ nuclear spin which can also suppress the line broadening 
caused by hydrogen nuclei, while keeping the interaction between $\ele{31}{P}$ and the NV spin once the Hartmann-Hann condition is matched. In our numerical simulation, we apply a magnetic field $B=290\mbox{G}$ such that the Larmor frequencies of $\ele{31}{P}$, $\ele{1}{H}$ and $\ele{13}{C}$ are quite different, namely equal to $500 \mbox{kHz}$, $1235 \mbox{kHz}$ and $310.6 \mbox{kHz}$ respectively. The driving amplitude of the NV spin is tuned to match the effective Larmor frequency of $\ele{31}{P}$, and in the mean time its interaction with the $\ele{13}{C}$ spin bath can be effectively suppressed (see appendix). The difference of Larmor frequencies also allow us to selectively drive the nuclei $\ele{1}{H}$ (e.g. with the driving amplitude $20 \mbox{kHz}$) while not affecting the other two types of nuclei. The distance between the nucleus $\ele{31}{P}$ and $\ele{1}{H}$ is about $0.2\mbox{nm}$, and the corresponding coupling constant is $g_m\simeq 6 \mbox{kHz}$. The line broadening due to the hydrogen nuclei is much larger than the coupling between  NV spin and the target nucleus $\ele{31}{P}$ (which is on the order of $1\mbox{kHz}$ at a distance of $5 \mbox{nm}$). We remark here that shallow (below $5\mbox{nm}$) implantation and isotropically engineered diamond has been realized experimentally \cite{Okai12,Ohno1207,Bala09,Pezz10}. In Fig.\ref{EP31}(a), it can be seen that if we do not drive the nuclei $\ele{1}{H}$, the interaction between $\ele{31}{P}$ and $\ele{1}{H}$ will smear the measured signal and we can hardly get information about the position of $\ele{31}{P}$. By applying a radio frequency driving field on resonance with the Larmor frequency of $\ele{1}{H}$, we can then clearly obtain information of the hyperfine vector (and thereby the directional vector) of $\ele{31}{P}$ with respect to NV spin, see Fig.\ref{EP31}(b). The distance between NV center and $\ele{31}{P}$ can be estimated by applying the magnetic field direction orthogonal to the hyperfine vector. The corresponding value of $J$ is $0.2041 \mbox{kHz}$, which agrees well with the exact value $0.2065 \mbox{kHz}$, see Fig.\ref{EP31}(c).

\begin{figure}[t]
\begin{center}
\hspace{-0.2cm}
\includegraphics[width=8cm]{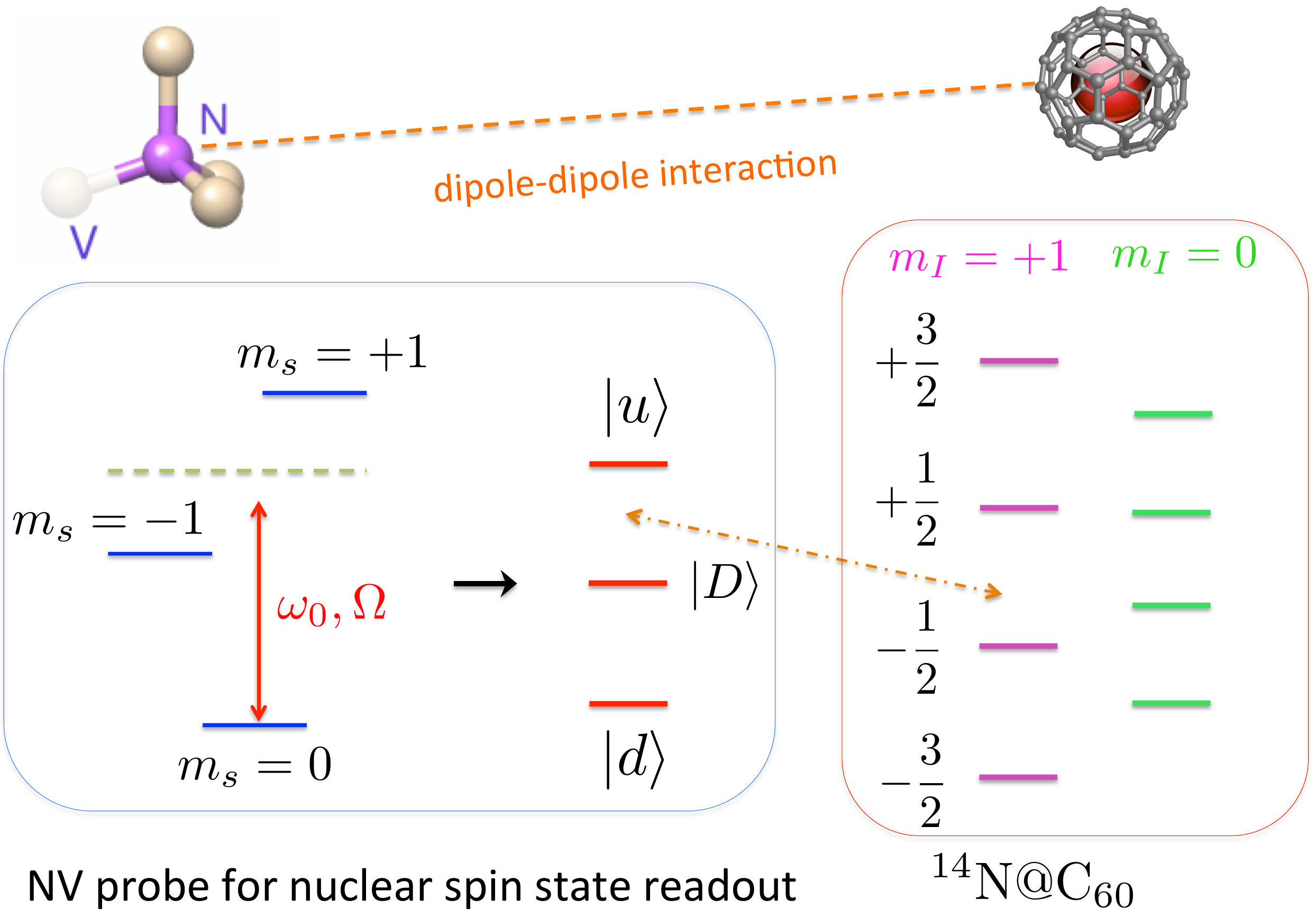}
\end{center}
\caption{(Color online) Scheme for quantum non-demolition measurement of the nitrogen nuclear spin state in $\ele{14}{N}@\mbox{C}_{60}$ with a NV center. For the nuclear spin state $\ket{0}\equiv \ket{m_I=0}$, the allowed electron transition frequency in  $\ele{14}{N}@\mbox{C}_{60}$ is $\omega_e$ and will not be on resonant with the NV dressed spin transition frequency $\omega_{NV}=(2\Omega^2+\omega_e^2)^{1/2}$ or $2(2\Omega^2+\omega_e^2)^{1/2}$, while for $\ket{1}\equiv \ket{m_I=+1}$, the resonance condition is satisfied when $\omega_e+a=\bla{2\Omega^2+\omega_e^2}^{1/2}$.}
\label{ENC60:setup}
\end{figure}
%

\begin{figure*}
\begin{center}
\hspace{-0.5cm}
\includegraphics[width=6cm]{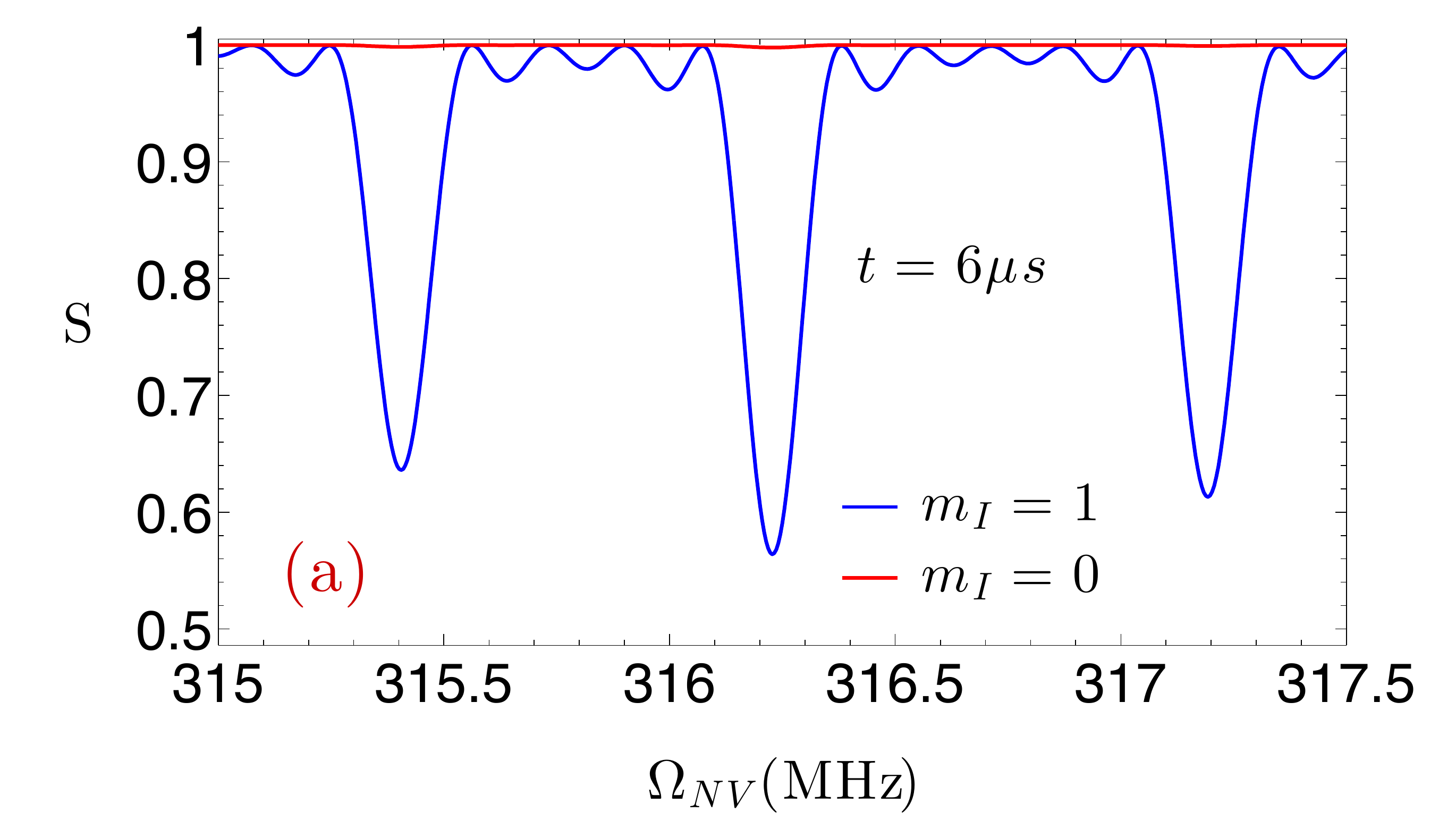}
\hspace{-0.2cm}
\includegraphics[width=6cm]{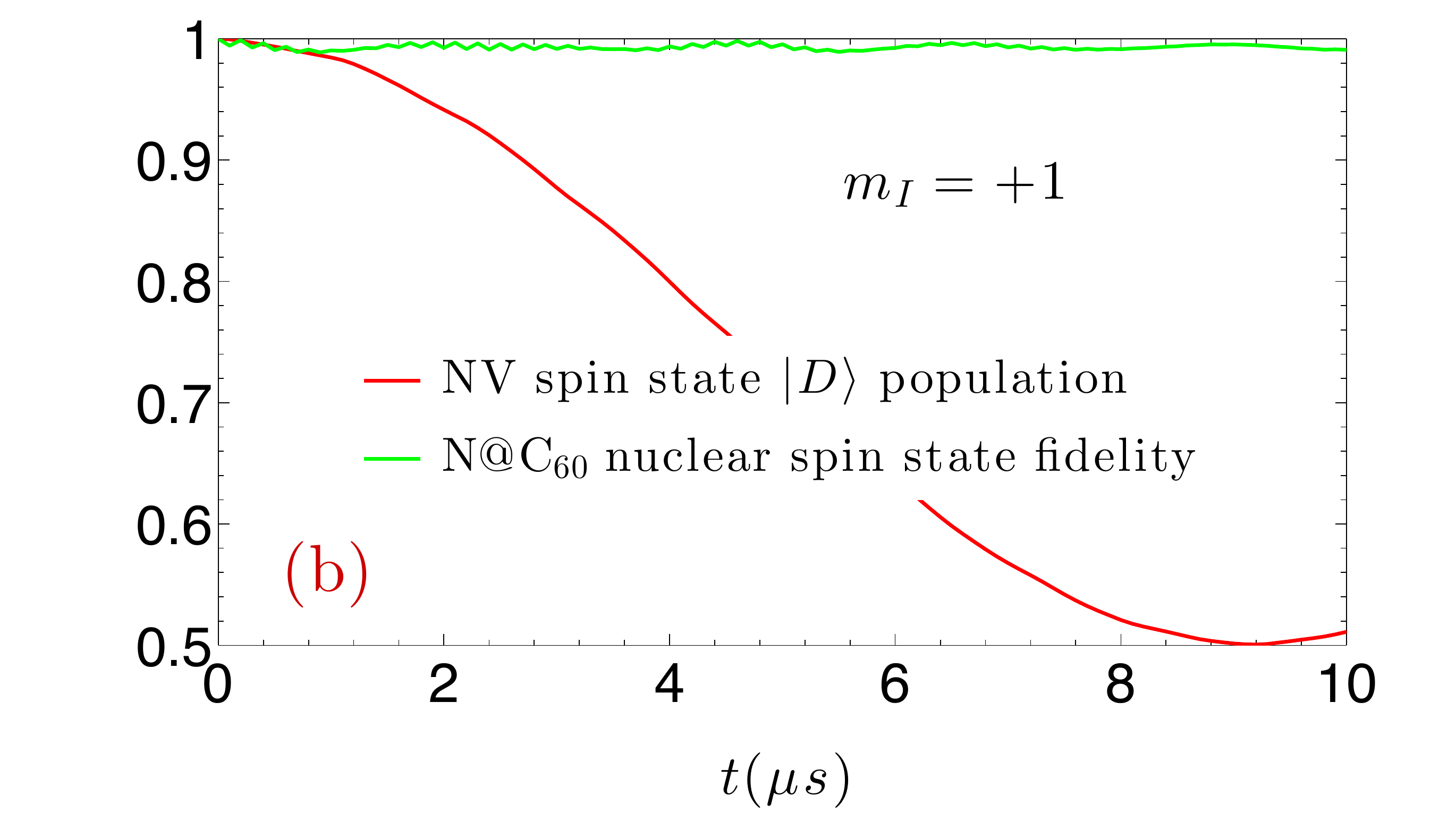}
\hspace{-0.2cm}
\includegraphics[width=6cm]{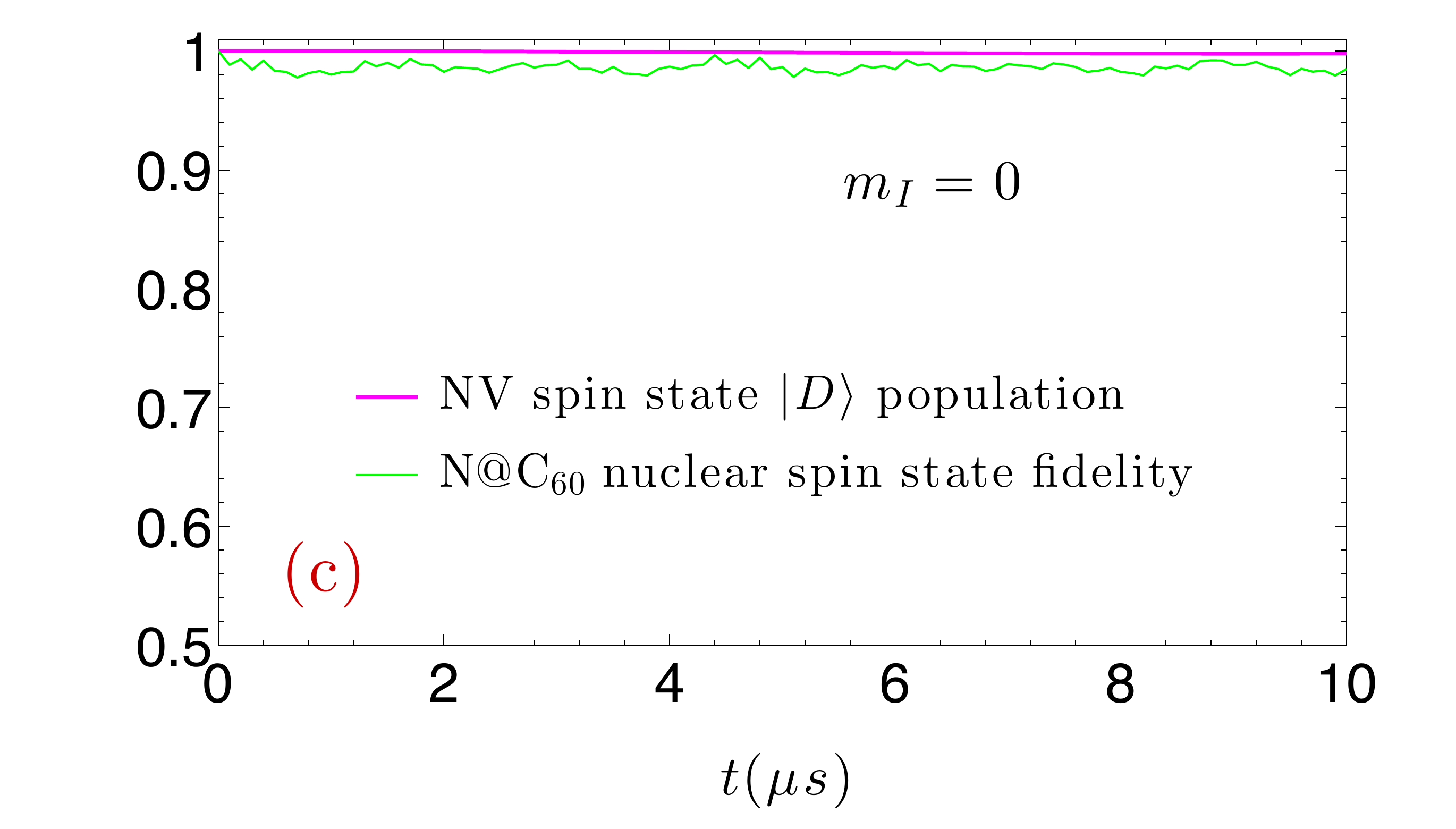}
\end{center}
\caption{(Color online) quantum non-demolition measurement of the nitrogen nuclear spin state in $\ele{14}{N}@\mbox{C}_{60}$ with a NV center at a distance of $8 \mbox{nm}$. (a) The signal $S$ (i.e. the state $\ket{D}$ population of NV spin) measured at time $t=6\mu s$ as function of the effective NV transition frequency $\omega_{NV}=(2\Omega^2+\omega_e^2)^{1/2}$ for two nuclear spin state $\ket{m_I=0}$ (red) and $\ket{m_I=+1}$ (blue). (b-c) The signal $S$ and the nuclear spin state fidelity as a function of time $t$ for the nuclear spin state $\ket{m_I=+1}$ (b) and $\ket{m_I=0}$ (c). The Rabi frequency $\Omega$ is chosen as $\Omega=70.7\mbox{MHz}$ corresponding to the resonant frequency $\Omega_{NV}=316.2 \mbox{MHz}$ for the nuclear spin state $\ket{m_I=+1}$. The applied magnetic field is $\gamma_e B =300 \mbox{MHz}$ with $B=107 \mbox{G}$. The electron spin of $\ele{14}{N}@\mbox{C}_{60}$ is not polarized, and its initial state is approximated by the maximally mixed state.}\label{ENC60}
\end{figure*}

\section{Measurement of single nuclear spin state} 
\label{spinstate:sec}

In our model of NV spectrometer, the flip-flop process happens under the Hartmann-Hahn resonance condition but also requires that the nuclear spin state is opposite to the NV dressed spin state. As nuclear spins have long coherence times and can serve as robust qubits, here we show that it is possible to perform (quantum non-demolition) measurement on the nuclear spin state of $\ele{14}{N}@\mbox{C}_{60}$ (i.e. a nitrogen atom in a $\mbox{C}_{60}$ cage) \cite{Twa02,Ben06} as an example to demonstrate our idea. The $\ele{14}{N}@\mbox{C}_{60}$ molecule has an electron spin-$\frac{3}{2}$ coupled to the $\ele{14}{N}$ nuclear spin-$1$. The hyperfine interaction is isotropic and the spin Hamiltonian is given by
\begin{equation}
H_S=\omega_e S_N^z+\omega_N I_N^z +  \Delta_Q \bla{I_N^z}^2 + a {\bf S}_N\cdot {\bf I}_N 
\end{equation}
where $\omega_e=-\gamma_e B$, $\omega_N=\gamma_N B$ and the quadrupole splitting is $\Delta_Q=5.1\mbox{MHz}$, the hyperfine coupling is $a=15.88 \mbox{MHz}$ \cite{Twa02,Ben06}. We can encode a qubit in the nuclear spin state $\ket{0}_I\equiv \ket{m_I=0}$ and $\ket{1}_I\equiv \ket{m_I=+1}$. If the difference between the electron and nuclear Zeeman splitting $\omega_e-\omega_N$ is large enough, the non-secular terms in the hyperfine coupling can be neglected, and thus the system Hamiltonian can be rewritten as follows
\begin{equation}
H_S=\omega_e S_N^z+\omega_N I_N^z +  \Delta_Q \bla{I_N^z}^2  + a {\bf S}_N^z {\bf I}_N^z.
\end{equation}
When the nuclear spin is in the state of $\ket{m_I=0}$, the electron spin energies is equidistant with $\omega_e$, while for the nuclear spin state $\ket{m_I=+1}$, it is $\omega_e+a$, see Fig.\ref{ENC60:setup}. By applying an additional magnetic field, the energy separation between the NV spin state $\ket{ms=+1}$ and  $\ket{ms=-1}$ is $\Delta=2\omega_e$, namely
\begin{equation}
H_{NV}=\bla{\omega_0-\omega_e}\ketbra{-1}{-1}+\bla{\omega_0+\omega_e}\ketbra{+1}{+1}.
\end{equation}
with $\omega_0=2.87\mbox{GHz}$. This means that we will inevitably drive both NV electronic spin transitions $\ket{0} \leftrightarrow \ket{\pm 1}$ if the driving amplitude $\Omega$ is around $\omega_e$ in order to satisfy the Hartmann-Hahn resonant condition between NV dressed spin and the electron spin in $^{14} \mbox{N}@\mbox{C}_{60}$, see section \ref{section:basic_model}. Thus, we need to take all three sublevels of the NV spin into account. Our idea is to apply a continuous driving field at frequency $\omega_0$ as
\begin{equation}
H_d=\Omega \cos(\omega_0 t) \bla{\ketbra{+1}{0}+\ketbra{-1}{0}+\mbox{h.c.}}.
\end{equation}
The driving field is off-resonant with the NV spin transitions with the detuning $\pm \omega_e$. In the interaction picture with respect to $H_{NV}$, we have 
\begin{equation}
H_I=(2\Omega^2+\omega_e^2)^{1/2}\bla{\ketbra{u}{u}-\ketbra{d}{d} },
\end{equation}
where we have three dressed states as 
\begin{eqnarray}
\ket{u}&=&\frac{1}{\eta_+^2+2} \bla{\eta_+^2 \ket{-1}+2\eta_+ \ket{0} +\ket{+1}},\\
\ket{D}&=&\frac{1} {(2\Omega^2+\omega_e^2)^{1/2}} \bla{\Omega \ket{+1}+ \omega_e \ket{0}-\Omega \ket{-1}},\\
\ket{d}&=&\frac{1}{\eta_-^2+2} \bla{\eta_-^2 \ket{-1}-2\eta_- \ket{0} +\ket{+1}},
\end{eqnarray}
where $\eta_{\pm}=\blb{(2\Omega^2+\omega_e^2)^{1/2} \pm\omega_e}/\Omega$. We note that the NV spin coupling operator ${\bf S}_z$  with the other spins will induce transitions among the dressed states with two transition frequencies $\omega_1=(2\Omega^2+\omega_e^2)^{1/2}$ and $\omega_2=2 (2\Omega^2+\omega_e^2)^{1/2}$, see Fig.\ref{ENC60:setup}. Therefore, we first prepare NV spin in the state $\ket{D}$ and then tune the Rabi frequency to be on resonance with the allowed electron transition frequency in the system $^{14} \mbox{N}@\mbox{C}_{60}$ corresponding to the nitrogen nuclear spin state $\ket{m_I=+1}$, namely to satisfy the following condition
\begin{equation}
(2\Omega^2+\omega_e^2)^{1/2}=\omega_e+a.
\end{equation}
After time $t$, we measure the probability that NV spin remains in the state $\ket{D}$. If the nuclear spin is in the state $\ket{m_I=0}$, it can be seen that the resonant condition would never be satisfied once we drive the NV spin (i.e. with $\Omega > 0$), the NV spin will thus stays in the initial state $\ket{D}$; otherwise if the nuclear spin state is $\ket{m_I=+1}$, the flip-flop process can happen and thus the state population of $\ket{D}$ will change once the resonant condition is satisfied. In Fig.\ref{ENC60}(a), we see that there are actually three resonant frequencies. The splitting comes from the virtual electronic transition caused by the nuclei ($\sim a^2/\omega_e= 0.84\mbox{MHz}$). In the mean time, the nuclear spin state populations are not affected by the readout procedure, and one can thus realize repetitive measurements of the nitrogen nuclear spin state.  Thus the readout represents a non-demolition measurement on the nuclear spin state that can be repeated. The feasibility of this idea is verified by our numerical simulation, see Fig.\ref{ENC60}(b-c). In our numerical simulation, we apply a magnetic field $|\gamma_e B|=300\mbox{MHz}$ (namely $B=107 \mbox{G}$), and the corresponding Larmor frequencies of $\ele{13}{C}$ and $\ele{14}{N}$ are $114.75 \mbox{kHz}$, $32.97 \mbox{kHz}$ respectively. The amplitude of driving on the NV spin corresponding to the central resonant frequency $\omega_{NV}=316.2\mbox{MHz}$ is $\Omega=70.7\mbox{MHz}$, which is strong enough to suppress the effect of the $\ele{13}{C}$ spin bath. It can be seen that the required time for one readout is much shorter than the coherence time $T_2$ of the electron spin in $^{14} \mbox{N}@\mbox{C}_{60}$ (which is $20\mu s$ at room temperature \cite{Twa02}) for a distance of $8 \mbox{nm}$ from $^{14}\mbox{N}@\mbox{C}_{60}$ to NV center. We remark that if we first polarize the electron spin to the state $\ket{-\frac{3}{2}}$ (e.g. by using NV center), it is possible to improve the readout efficiency. We also would like to point out that the present mechanism of measuring nuclear spin state implies that one can transfer the polarization of the NV spin to the other electron/nuclear spins, which can be exploited to achieve dynamical spin polarization. 

\section{Measure spin-spin distance and alignment} 
\label{spinpair:sec}

\begin{figure}[b]
\begin{center}
\includegraphics[width=8cm]{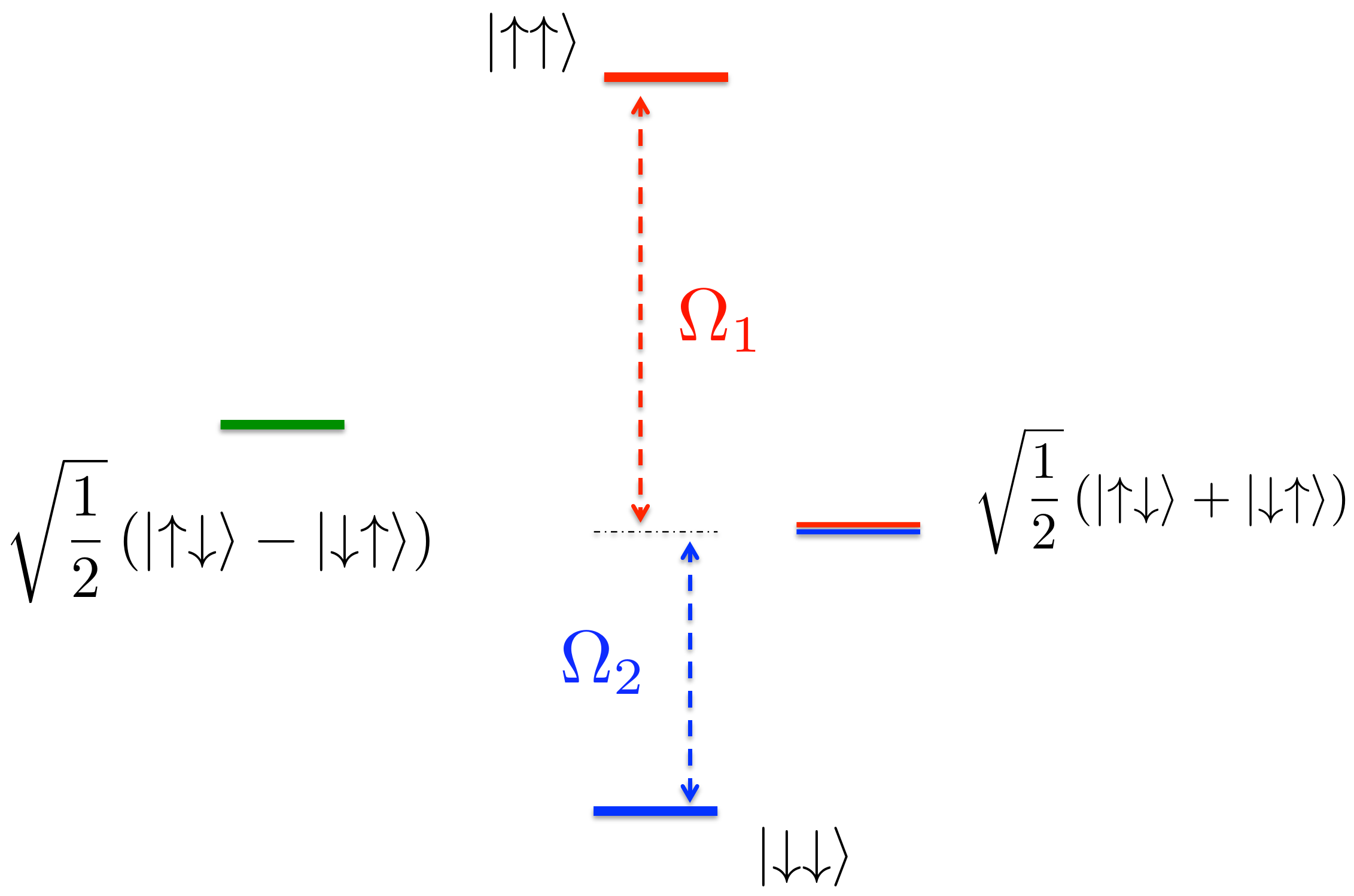}
\end{center}
\caption{(Color online) Eigenstates of two interacting spins under a strong magnetic field. We use NV spin to detect the transition frequencies of $\Omega_1$ and $\Omega_2$, the difference between which provides the information about the spin-spin coupling strength and the alignment direction of the spin pair with respect to the applied magnetic field. }\label{MOSP}
\end{figure}
%

%
\begin{figure*}
\begin{center}
\begin{minipage}{20cm}
\hspace{-2.2cm}
\includegraphics[width=18cm]{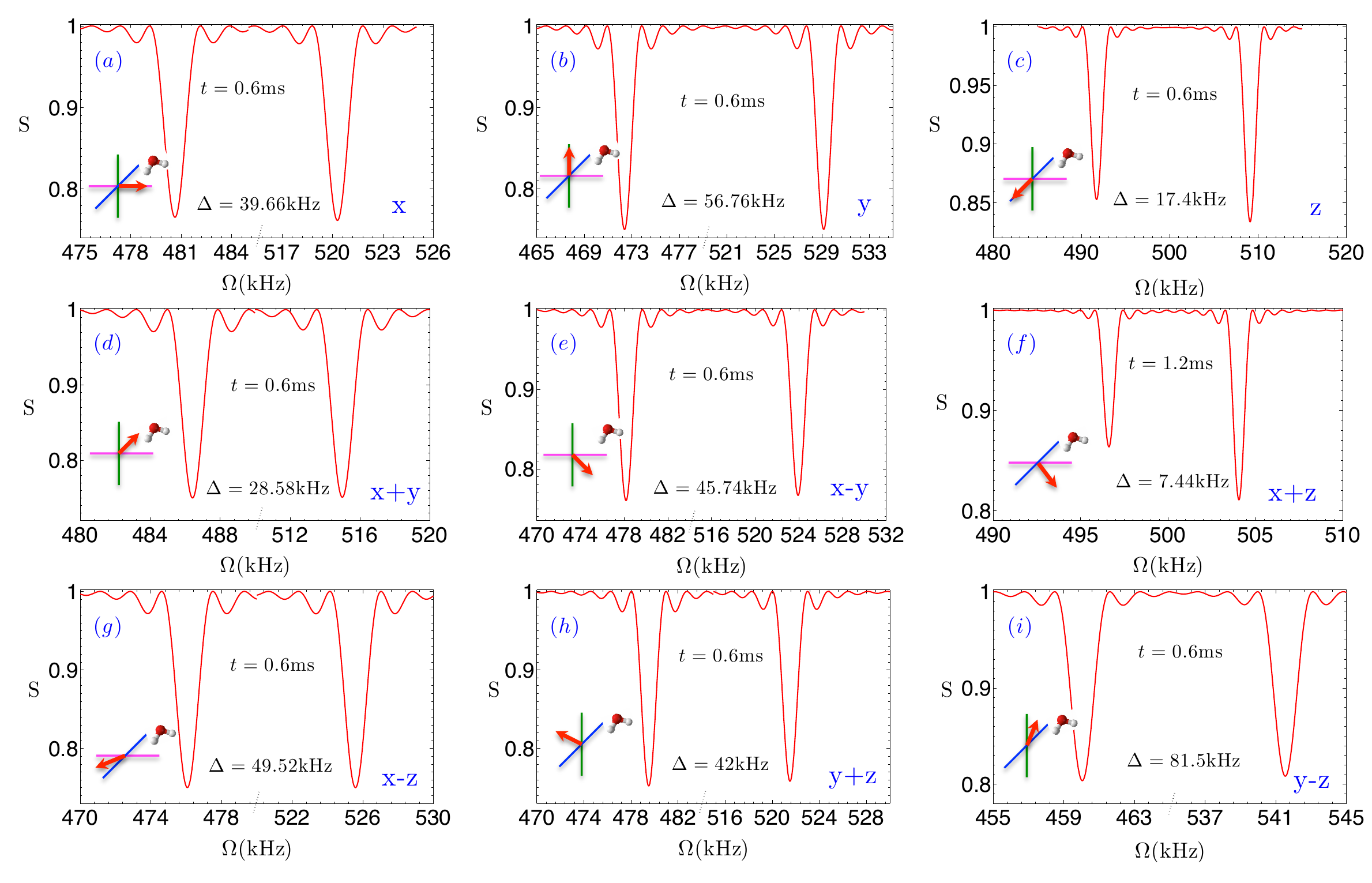}
\end{minipage}
\end{center}
\caption{(Color online) Resonance of a single $\ele{1}{H}_2\mbox{O}$ molecule with a NV center under an additional magnetic field $\gamma_{\ele{1}{H}} B=500\mbox{kHz}$ (namely $B=117 \mbox{G}$) in different directions: $\hat{x}$ (a),  $\hat{y}$ (b), $\hat{z}$ (c), $\hat{x}+\hat{y}$ (d), $\hat{x}-\hat{y}$ (e), $\hat{x}+\hat{z}$ (f), $\hat{x}-\hat{z}$ (g), $\hat{y}+\hat{z}$ (h), $\hat{y}-\hat{z}$ (i). The signal $S$ is measured at time $t$. In each plot, $\Delta$ denotes the distance between two resonant frequencies around the Larmor frequency as in Eq.(\ref{Omega_1}-\ref{Delta_H2O}). The $\ele{1}{H}_2\mbox{O}$ molecule is $5\mbox{nm}$ from the NV center, and the vector that connects two hydrogen nuclei is described by $(r,\theta_0,\phi_0)=(0.1515 \mbox{nm},118.2^o,288.85^o)$. From these results, we can infer that the distance is $r=0.1518 \mbox{nm}$, and the alignment direction is $(\theta,\phi)=(118.29^o,288.82^o)$ following Eq.(\ref{rsquare:H2O-1}-\ref{rsquare:H2O-3}), which are in good agreement with the exact parameters.}\label{EH2O}
\end{figure*}
%

For the tracking of molecular conformations or the determination of molecular structure it can be of advantage to determine the distance between two specific external nuclei. We first consider a target system which consists of two (electron or nuclear) spins interacting with each other. The system Hamiltonian is written as
\begin{equation}
H_{S}= -\gamma_{N} {\bf B} \cdot \bla{ {\bf I}_{N_1} + {\bf I}_{N_2}} + g  \blb{{\bf I}_{N_1}\cdot {\bf I}_{N_2}-3\bla{{\bf I}_{N_1}\cdot \hat{r}}\bla{{\bf I}_{N_2}\cdot \hat{r}}}.
\end{equation}
The interaction strength $g=\bla{4\pi \hbar\mu_0 \gamma_{N}^2}/ r^3$ depends on the intra-distance $R$ between two spins. The  vector $\vec{r}=r\hat{r}$ connects two spins and the unit vector $\hat{r}$ characterizes the alignment of spin pair. To measure their intra-distance $r$ and the orientation of the alignment vector $\hat{r}$, we apply a strong magnetic field $\omega_N=\gamma_N B \gg g$, as the energy spectrum will depend on the relative orientation between the alignment vector $\hat{r}$ and the applied magnetic field. On the other hand, the NV spectrometer will thus be operating with a strong enough driving field to sufficiently decouple from the other external noise. Under such a condition, the system Hamiltonian can be rewritten as follows
\begin{equation}
H_S=\omega_N \bla{I_{N_{1}}^z+I_{N_{2}}^z}+g_{12}\blb{ I_{N_{1}}^z I_{N_{2}}^z-\frac{1}{2} \bla{I_{N_{1}}^x I_{N_{2}}^x+I_{N_{1}}^y I_{N_{2}}^y } },
\end{equation}
where $g_{12}=g\bla{1-3 \cos^2\theta}$, with $\cos\theta=\hat{r}\cdot \hat{b}(\theta,\phi)$, and the spin operators $I^x, I^y, I^z$ are defined in the quantization axis induced by the applied magnetic field. The eigenstates and the corresponding energies can be written as
\begin{eqnarray}
\ket{E_0}&=&\ket{\uparrow \uparrow}, E_0=\omega_N+\frac{g}{4} \bla{1-3 \cos^2\theta} ,\\
\ket{E_1}&=&\sqrt{\frac{1}{2}} \bla{ \ket{\uparrow \downarrow} + \ket{\downarrow \uparrow} }, E_1=-\frac{g}{2} \bla{1-3\cos^2\theta},\\
\ket{E_2}&=&\sqrt{\frac{1}{2}} \bla{ \ket{\uparrow \downarrow} - \ket{\downarrow \uparrow} }, E_2=0,\\
\ket{E_3}&=&\ket{\downarrow \downarrow}, E_3=-\omega_N+\frac{g}{4} \bla{1-3 \cos^2\theta} .
\end{eqnarray}
The transition from the singlet eigen state $\ket{E_2}$ to the states $\ket{E_0}$ and $\ket{E_3}$ is determined by the inhomogeneity of the coupling operators of two spins to the NV center, namely $\hat{A}_1-\hat{A}_2$, which is usually small if the intra spin distance is much smaller than their distance from NV center. Therefore, if we tune the Rabi frequency of the NV spin around the Zeeman energy $\omega_N$, the dominant flip-flop processes mainly happen with the transitions from $\ket{E_1}$ to $\ket{E_0}$ and $\ket{E_3}$, see Fig.\ref{MOSP}, with the following two corresponding resonant frequencies
\begin{eqnarray}
\Omega_1&=&\omega_N+\frac{3g}{4} \bla{1-\cos^2\theta},\label{Omega_1}\\
\Omega_2&=&\omega_N-\frac{3g}{4} \bla{1-\cos^2\theta}.\label{Omega_2}
\end{eqnarray}
The difference between these two resonant frequencies is 
\begin{equation}
\Delta=\frac{3g}{2} |1-3 \cos^2\theta|,
\label{Delta_H2O}
\end{equation}
which provides information about the coupling strength $g$ and the alignment vector $\hat{r}$. To determine their exact values, we propose to apply magnetic fields in different (nine) different directions, and measure the resonant frequencies respectively as follows
\begin{eqnarray}
\Delta_x&=&\frac{3g}{2} |1-3 r_x^2|, \label{Deltax}\\
\Delta_y&=&\frac{3g}{2} |1-3 r_y^2|,\\
\Delta_z&=&\frac{3g}{2} |1-3 r_z^2|,\\
\Delta_{x+y}&=&\frac{3g}{2} |1-\frac{3}{2} (r_x^2+r_y^2+2r_xr_y)|,\label{Deltaxpy}\\
\Delta_{x-y}&=&\frac{3g}{2} |1-\frac{3}{2} (r_x^2+r_y^2-2r_xr_y)|,\\
\Delta_{x+z}&=&\frac{3g}{2} |1-\frac{3}{2} (r_x^2+r_z^2+2r_xr_z)|,\\
\Delta_{x-z}&=&\frac{3g}{2} |1-\frac{3}{2} (r_x^2+r_z^2-2r_xr_z)|,\\
\Delta_{y+z}&=&\frac{3g}{2} |1-\frac{3}{2} (r_y^2+r_z^2+2r_y r_z)|,\\
\Delta_{y-z}&=&\frac{3g}{2} |1-\frac{3}{2} (r_y^2+r_z^2-2r_y r_z)|.\label{Deltaymz}
\end{eqnarray}
After some calculations, we can obtain
\begin{eqnarray}
\blb{\bla{\frac{3g}{2} } r_x r_y}^2&=&\frac {\bla{\Delta_{x-y}^2-\Delta_{x+y}^2}^2}  {36\Delta_z^2},\\
\blb{\bla{\frac{3g}{2} } r_x r_z}^2&=&\frac {\bla{\Delta_{x-z}^2-\Delta_{x+z}^2}^2}  {36\Delta_y^2},\\
\blb{\bla{\frac{3g}{2} } r_y r_z}^2&=&\frac {\bla{\Delta_{y-z}^2-\Delta_{y+z}^2}^2}  {36\Delta_x^2}.\label{ryz}
\end{eqnarray}
From Eq.(\ref{Deltax}-\ref{ryz}), we calculate the coupling strength $g$ as
\bea
\nonumber
g^2=&&\frac{2}{27} \bla{\Delta_x^2+\Delta_y^2+\Delta_z^2}+\frac{1}{27}\blb{\frac {\bla{\Delta_{x-y}^2-\Delta_{x+y}^2}^2}  {\Delta_z^2}}\\
&&+\frac{1}{27}\blb{\frac {\bla{\Delta_{y-z}^2-\Delta_{y+z}^2}^2}  {\Delta_x^2}+\frac {\bla{\Delta_{x-z}^2-\Delta_{x+z}^2}^2}  {\Delta_y^2}}.\label{gsquare:H2O}
\eea
Furthermore, we get
\begin{equation}
r_x^2=\frac{1}{3}-\frac{1}{27g^2} \blb{4\Delta_x^2+\frac {\bla{\Delta_{x-y}^2-\Delta_{x+y}^2}^2}  {\Delta_z^2}+\frac {\bla{\Delta_{x-z}^2-\Delta_{x+z}^2}^2}  {\Delta_y^2}},\label{rsquare:H2O-1}
\end{equation}
\begin{equation}
r_y^2=\frac{1}{3}-\frac{1}{27g^2} \blb{4 \Delta_y^2+\frac {\bla{\Delta_{x-y}^2-\Delta_{x+y}^2}^2}  {\Delta_x^2}+\frac {\bla{\Delta_{y-z}^2-\Delta_{y+z}^2}^2}  {\Delta_z^2}},\label{rsquare:H2O-2}
\end{equation}
\begin{equation}
r_z^2=\frac{1}{3}-\frac{1}{27g^2} \blb{4 \Delta_z^2+\frac {\bla{\Delta_{y-z}^2-\Delta_{y+z}^2}^2}  {\Delta_x^2}+\frac {\bla{\Delta_{x-z}^2-\Delta_{x+z}^2}^2}  {\Delta_y^2}}.\label{rsquare:H2O-3}
\end{equation}
Finally, with the obtained values of $g^2,r_x^2,r_y^2,r_z^2$ we compare Eq.(\ref{Deltaxpy}-\ref{ryz}) with the above equations and can then decide the relative signs between $r_x,r_y,r_z$ and thereby derive the alignment vector $\hat{r}$ of the spin pair.

\subsection{Measure distance between two hydrogen nuclei}

To demonstrate the basic principles, we have applied our ideas to the simple example of measuring the distance and alignment of two hydrogen nuclei in a water molecule, e.g. lying on diamond surface. In Fig.\ref{EH2O}, we plot the resonance frequencies with a magnetic field along different directions as in Eqs.(\ref{Deltax}-\ref{Deltaymz}). In our numerical simulation, the strength of the magnetic field is such that $\gamma_{\ele{1}{H}} B=500\mbox{kHz}$ (namely $B=117 \mbox{G}$). The amplitude of driving on the NV spin is thus strong enough to suppress the effect of the $\ele{13}{C}$ spin bath in diamond. The $\ele{1}{H}_2\mbox{O}$ molecule is assumed to be $5\mbox{nm}$ from the NV center, and the vector that connects two hydrogen atoms is described by $(r,\theta_0,\phi_0)= (0.1515 \mbox{nm},118.2^o,288.85^o)$. With the calculated resonant frequencies as shown in Fig.\ref{EH2O}, we follow the equations in Eq.(\ref{gsquare:H2O}-\ref{rsquare:H2O-3}) and obtain that $g=34.684 \mbox{kHz}$, and thereby we infer that the intra-molecular distance is $d=0.1518 \mbox{nm}$, and the alignment direction is $(\theta,\phi)=(118.29^o,288.82^o)$, which are in good agreement with the exact parameters of a water molecule. We remark that for different magnetic field directions, the effective flip-flop rate between the NV spin and the target system may be different. Thus, in Fig.\ref{EH2O} (f), we choose a longer duration time to demonstrate a resonant dip with a depth comparable with the other magnetic directions.

\subsection{Measure distance between two organic spin labels}

\label{spinlabel:sec}

\begin{figure}[t]
\begin{center}
\begin{minipage}{9cm}
\hspace{-0.6cm}
\includegraphics[width=4.6cm]{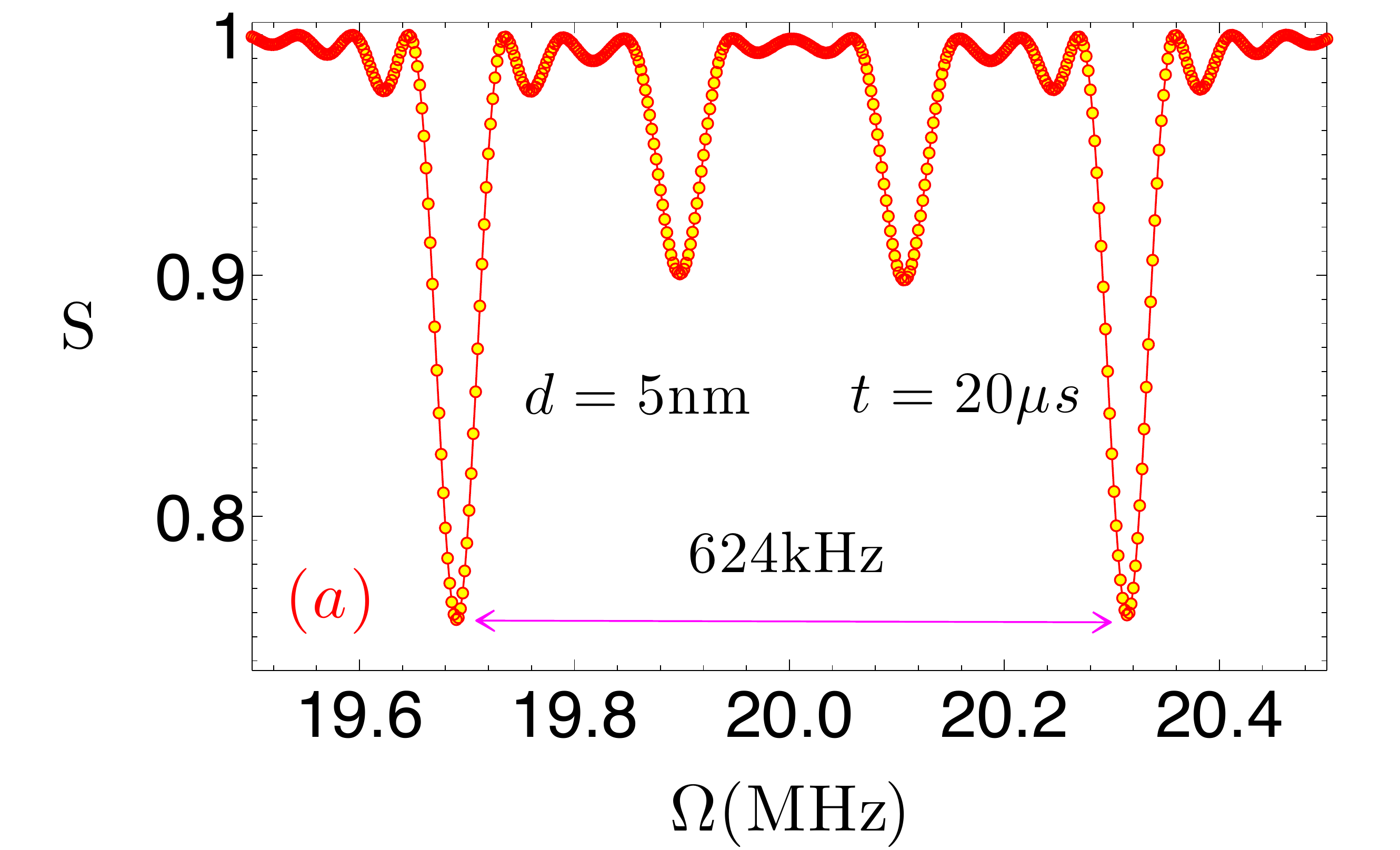}
\hspace{-0.3cm}
\includegraphics[width=4.6cm]{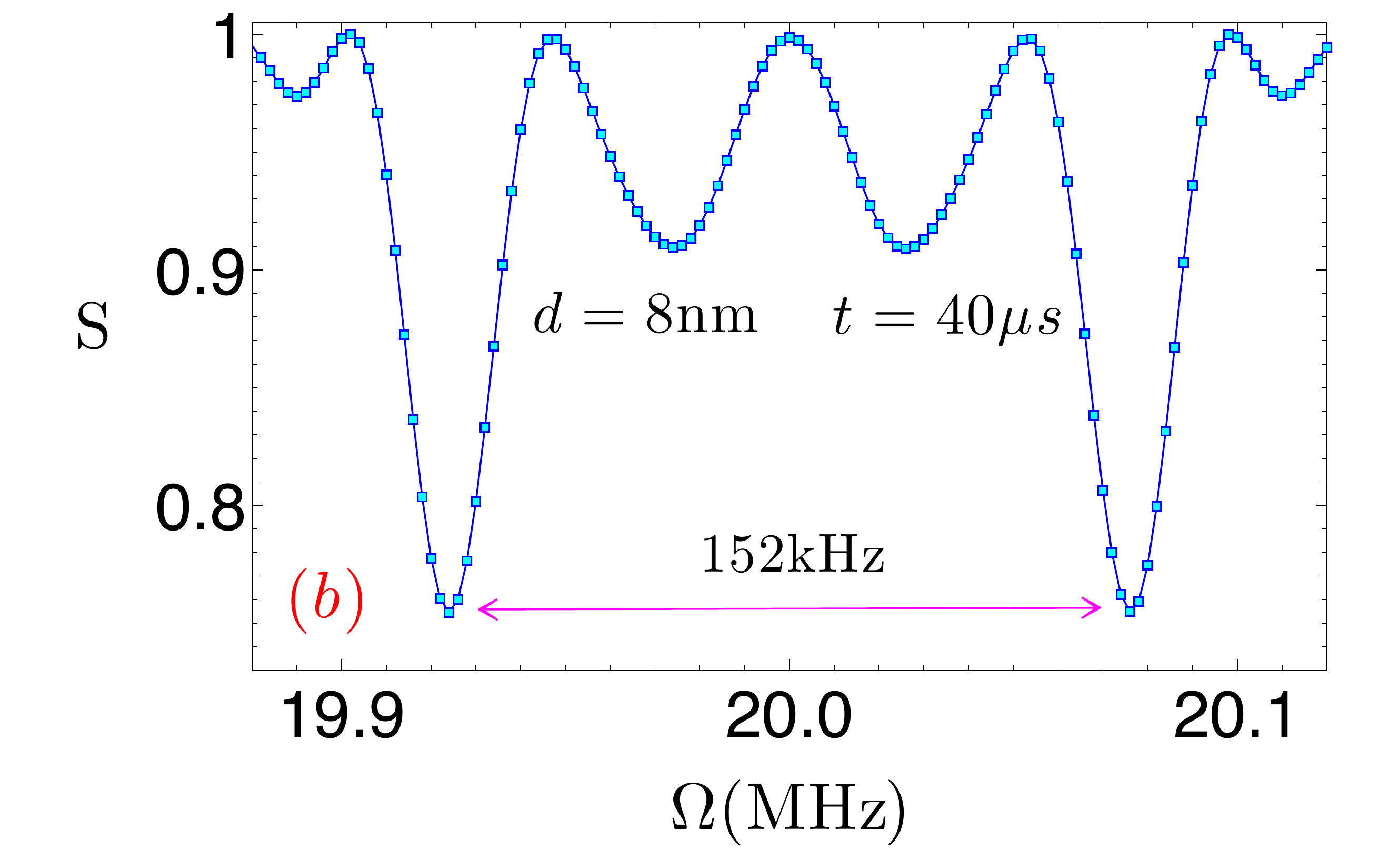}
\end{minipage}
\end{center}
\caption{(Color online)  Resonance of a pair of nitroxide spin labels with a distance of $d$. The magnetic field along the vector which connects two spin labels. Continuous driving field with Rabi frequency of $20\mbox{MHz}$ is applied on the spin labels. (a) The distance is $d=5 \mbox{nm}$, the coupling strength is $g=416.1\mbox{kHz}$, which corresponds to the difference between two resonant frequencies $\Delta_1=\frac{3}{2}g=624.1 \mbox{kHz}$. The signal is measured at time $t=20\mu s$. (b) The distance is $d=8 \mbox{nm}$, the coupling strength is $g=101.6\mbox{kHz}$, which corresponds to the difference between two resonant frequencies $\Delta_1=\frac{3}{2}g=152.4 \mbox{kHz}$. The signal is measured at time $t=40\mu s$.}\label{SP2}
\end{figure}

The protocol can be combined with spin labels and have potential applications in chemistry and biology ranging from determine biological structure and monitoring macro-molecule motions. Spin labels are organic molecules with a stable unpaired electron \cite{SPREFMT}. They can be attached to the protein (covalent or as a ligand) via a functional group. Electron spin resonance (ESR) on an ensemble based on spin labels has widely been used as a spectroscopic ruler to determine protein structure and monitor macromolecular assembly processes. However, it is very hard to go beyond a distance of $5 \mbox{nm}$ between spin labels, because e.g. inhomogeneous line broadening limits the spatial resolution \cite{SPREFMT}. We consider the widely used nitroxide spin labels and show that the resonance linewidth (which mainly depends on the coupling strength between NV spin and spin labels) assisted by dynamical nuclear polarization and continuous drivings can be narrow enough to resolve the resonance frequency splitting for a pair of spin labels with a distances larger than $5 \mbox{nm}$, see Fig.\ref{SP2}(b) for an example of $8 \mbox{nm}$. To suppress the effect of the nitrogen nuclear spin and enhance the signal, one can use the NV spin to first polarize the electron spin and then use the polarized electron spin to prepare the nuclear spin into the state $\ket{m_I=0}$. The coupling with the other nuclear spins is weak and can be suppressed by continuously driving the electron spin. If the driving amplitude is much stronger than the hyperfine coupling, the effective Hamiltonian is written as
\begin{eqnarray}
H_{NV-S}=&&\Omega_0 {\bf S}_x +\Omega \bla{S_{1}^x+S_{1}^x}+{\bf S}_z \bla{A_{1} S_{1}^z+A_{2} S_{2}^z}\\\nonumber
&&+g\bla{1-3 \cos^2\theta} \blb{ S_{1}^z 
S_{2}^z-\frac{1}{2} \bla{S_{1}^x S_{2}^x+S_{1}^y S_{2}^y } },
\end{eqnarray}
where $\cos\theta=\hat{r}\cdot \hat{b}(\theta,\phi)$, ${\bf S}$ is the NV spin operator and $S_i$ is the spin label operator, and the quantization axis of the spin labels are induced by the applied magnetic field. In the case that $\Omega \gg g$, the Hamiltonian for the spin labels can be approximated as follows
\begin{eqnarray}
H_{S}=&&\Omega \bla{S_{1}^x+S_{1}^x}-\frac{g}{2} \bla{1-3 \cos^2\theta} S_{1}^x S_{2}^x\\\nonumber
&&+\frac{g}{8}\bla{1-3 \cos^2\theta}
\bla{ \ket{\uparrow}_x\bra{\downarrow}\otimes \ket{\downarrow}_x\bra{\uparrow}+h.c. }.
\label{EFH}
\end{eqnarray}
where $\ket{\uparrow}_x$ and $\ket{\downarrow}_x$ are the eigenstates of the spin operator $S^x_i$. The eigenstates are similar to Fig.\ref{MOSP}, and can be explicitly written as follows with the corresponding energies
\begin{eqnarray*}
\ket{E_0}&=&\ket{\uparrow _x\uparrow_x}, E_0=\Omega -\frac{3g}{8} \bla{1-3 \cos^2\theta},\\
\ket{E_1}&=&\sqrt{\frac{1}{2}} \bla{ \ket{\uparrow _x \downarrow_x} + \ket{\downarrow_x \uparrow_x} }, E_1=\frac{g}{4} \bla{1-3 \cos^2\theta},\\
\ket{E_2}&=&\sqrt{\frac{1}{2}} \bla{ \ket{\uparrow_x \downarrow_x} - \ket{\downarrow_x \uparrow_x} }, E_2=0,\\
\ket{E_3}&=&\ket{\downarrow_x \downarrow_x}, E_3=-\Omega-\frac{3g}{8} \bla{1-3 \cos^2\theta} .
\end{eqnarray*}
The transitions from the eigenstate $\ket{E_1}$ to the other two eigenstates $\ket{E_0}$ and $\ket{E_3}$ correspond to the resonant frequencies as follows
\begin{eqnarray}
\Omega_1&=&\Omega+\frac{3g}{8} \bla{1-3 \cos^2\theta},\\
\Omega_2&=&\Omega-\frac{3g}{8} \bla{1-3 \cos^2\theta},
\end{eqnarray}
and the difference between these two resonant frequencies is
\begin{equation}
\Delta_1=\frac{3g}{4} \vert 1-3\cos^2\theta\vert.
\end{equation}
When the distance between two spin labels is large, such that the coupling operators $A_1$ and $A_2$ are inhomogeneous, 
the transition from the singlet state $\ket{E_2}$ to the states $\ket{E_0}$ and $\ket{E_3}$ may also be observed (see the two less pronounced resonant peaks in Fig.\ref{SP2}) with the following resonant frequencies
\begin{eqnarray}
\Omega_3&=&\Omega+\frac{g}{8} \bla{1-3\cos^2\theta},\label{Omega_3}\\
\Omega_4&=&\Omega-\frac{g}{8} \bla{1-3 \cos^2\theta},
\end{eqnarray}
and the difference between these two resonant frequencies is
\begin{equation}
\Delta_2=\frac{g}{4} \vert 1-3\cos^2\theta\vert.
\end{equation}

\subsection{Monitor the charge recombination of radical pair}
\label{rpm:sec}

Many chemical reactions involves radical pair intermediate, which consists of two unpaired electrons \cite{Ste89}. The radical pair mechanism has raised new interest recently regarding the potential role of quantum effect in this model to explain avian magnetoreception \cite{Ritz00,Cai10prl,Gau11prl,Ritz11,Cai11prl,Cai12pra,Mour12}. The radicals are in a charge separated state and interact with each other via exchange and dipole interactions. For the distance larger than $1\mbox{nm}$, the exchange interaction is negligible, and the main contribution comes from the dipole interaction. Here, we consider a simple model radical pair reaction, namely the radical pair is created in the singlet state and recombine at the same rate $k$ for both singlet and triplet states, the dynamics of which can be described by the following master equation as
\begin{eqnarray}
\nonumber
\frac{d}{dt} \rho=&&-i[H,\rho]-\frac{1}{2} \bla{ L_S^{\dagger} L_S \rho  +  \rho L_S^{\dagger} L_S -2 L_S \rho L_S^{\dagger} }\\
&&-\frac{1}{2} \bla{ L_T^{\dagger} L_T \rho  +  \rho L_T^{\dagger} L_T -2 L_T \rho L_T^{\dagger} }
\end{eqnarray}
where $H$ is the system Hamiltonian that describes the interaction between two radicals and the coupling between NV spin and radicals when the radical pair is in the charge separated state, and the Lindblad operators $L_S$, $L_T$ describe the recombination of the singlet and triplet radical pair into the product state, which are written as
\begin{eqnarray}
L_S&=& k^{1/2} \bla{ Q_S \otimes \ketbra{P}{S}}\\
L_T&=& k^{1/2} \bla{ Q_T \otimes \ketbra{P}{S}}
\end{eqnarray}
\begin{figure}[t]
\begin{center}
\begin{minipage}{9cm}
\hspace{-0.5cm}
\includegraphics[width=4.5cm]{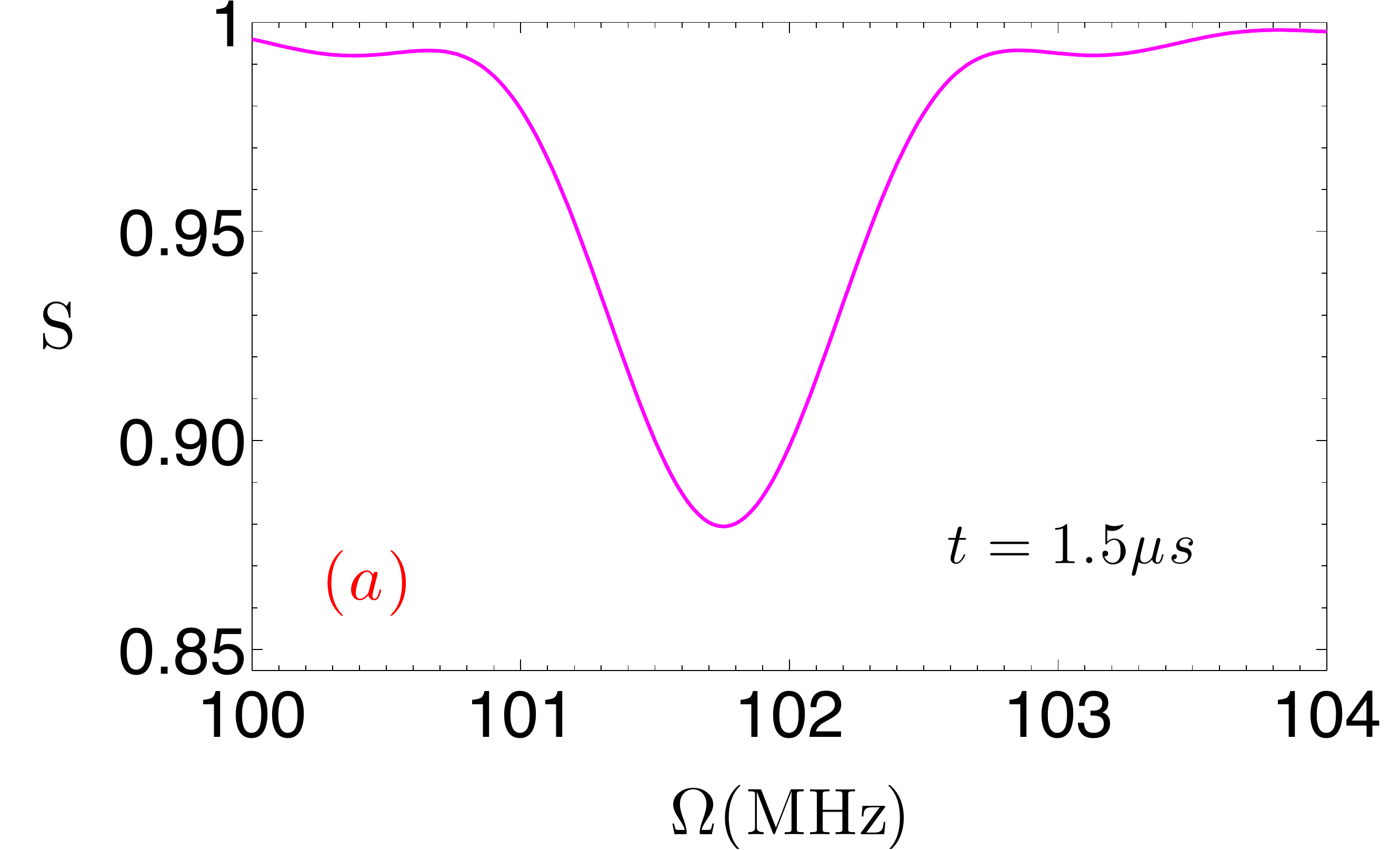}
\hspace{-0.3cm}
\includegraphics[width=4.5cm]{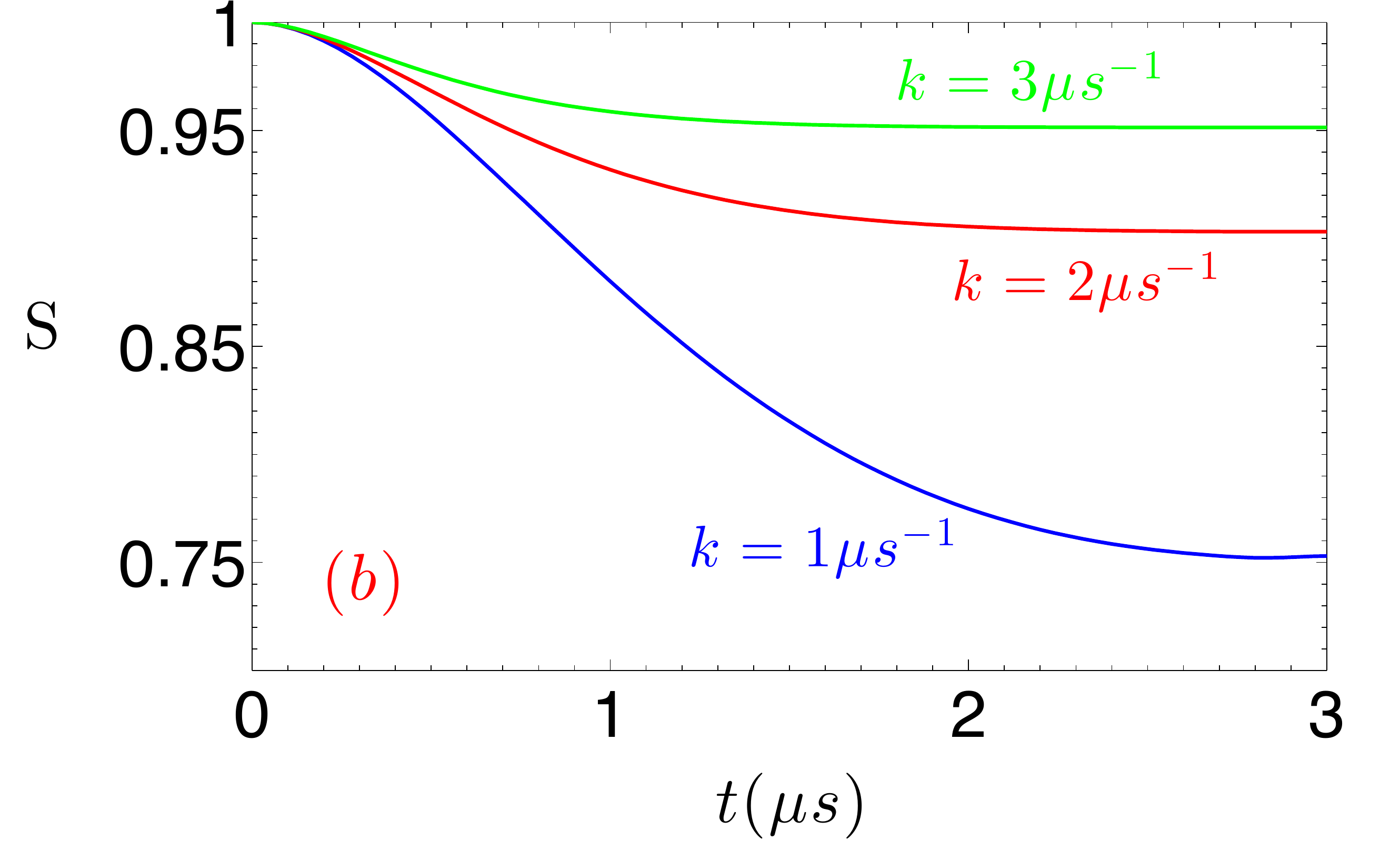}
\end{minipage}
\end{center}
\caption{(Color online)  (a) Resonance of a model  radical pair reaction with a distance of $d=2 \mbox{nm}$ between two radicals, the recombination rate is $k=1\mu s^{-1}$. The magnetic field is assumed to be along the vector which connects two radicals. The results are similar for other magnetic field directions. Continuous driving field with Rabi frequency of $100\mbox{MHz}$ is applied on the radical pair. (b) The Rabi frequency of NV spin is set at the resonant frequency shown in (a). As the radical pair recombines, the effective flip-flop rate decreases and the slope of the signal $S$ becomes smooth. }\label{RPMSI}
\end{figure}
with $Q_S$ and $Q_T$ the projectors into the singlet and triplet subspace, $\ket{S}$ and $\ket{P}$ represent the charge separate state and product state of radial pair respectively. The above master equation is equivalent to the conventional Haberkorn approach \cite{HAB}. We apply the same idea as in the model of spin labels (see section \ref{spinlabel:sec}) by applying an additional magnetic field and also continuously driving the radical spins to suppress the effect of surrounding nuclei. We assume the radical pair is created in the singlet state, so the resonant frequency near the driving Rabi frequency $\Omega$ is 
\begin{equation}
\Omega_3=\Omega+\frac{g}{8} \bla{1-3\cos^2\theta},
\end{equation}
with $\cos\theta=\hat{r}\cdot \hat{b}(\theta,\phi)$ (the same as Eq.(\ref{Omega_3})).The charge recombination leads to the decay of the effective flip-flop rate. If the recombination rate is comparable or smaller than the coupling between NV spin and the radicals, it is possible to observe the resonance frequency, see an example in Fig.\ref{RPMSI} (a). Therefore, if we tune the Rabi frequency of NV spin at the resonant frequency, we can monitor the recombination of the radical pairs by observing the decay of the flip-flop rate. In other words, it can serve as an evidence for the charge recombination. We remark that this provides possibility to see chemical reaction process at a single molecule level and may give insights into how radicals recombine into the product states.

\section{Summary and outlook} 

We have proposed a scheme to construct a nano-scale single molecule spectrometer based on NV centers in diamond under continuous driving. This spectrometer is tunable by changing the Rabi frequency. We demonstrate its application in the detection of a single nucleus, including its position and spin state. The idea can also be used to measure the distance and alignment of a spin pair. This opens a novel route to determine the structure of proteins and monitor conformational changes and other processes of relevance to biology, chemistry and medicine. We expect that our result and its extension can greatly enrich the diamond-based quantum technologies and their applications in chemistry and biology. The implementation of the present proposal of single molecule spectroscopy will benefit from the experimental developments including: shallow implanted NV centers in diamond \cite{Okai12,Ohno1207}, stable microwave driving fields \cite{Cai11}  and static magnetic field. A practical challenge when considering the biological applications is to localize the biomolecules properly on diamond surface and close to the NV centers. A theoretical challenge would be to calculate the resolution of the proposed microscope. Meaning, if we have a few atoms of the same type the efficiency of resolving them would be an interesting theoretical problem in Hamiltonian estimation which we intend to study.

\section{Appendix}

\subsection{Continuous dynamical decoupling of the $\ele{13}{C}$ spin bath} 

\begin{figure}[t]
\begin{center}
\begin{minipage}{9cm}
\hspace{-0.3cm}
\includegraphics[width=4.6cm]{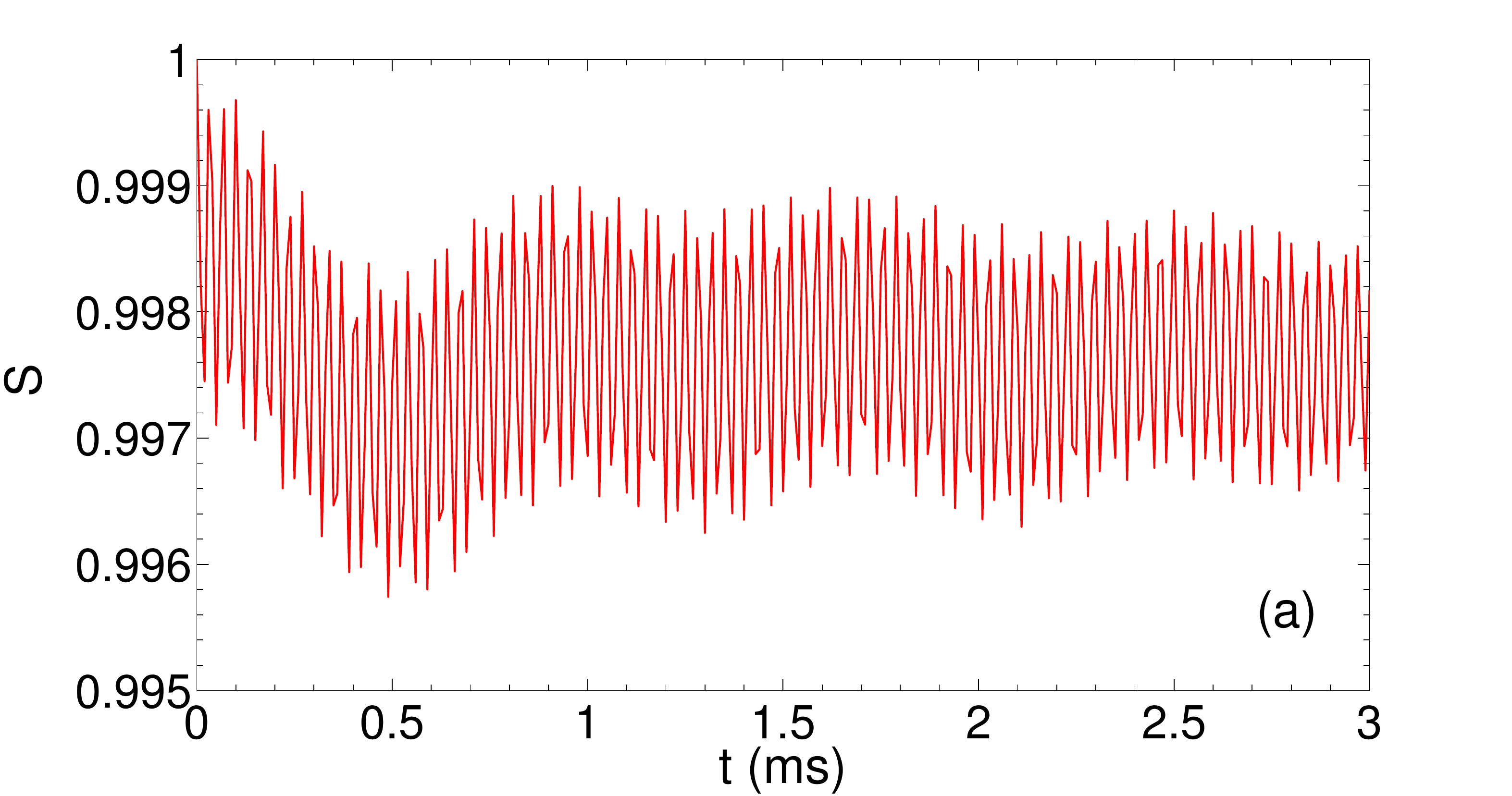}
\hspace{-0.3cm}
\includegraphics[width=4.6cm]{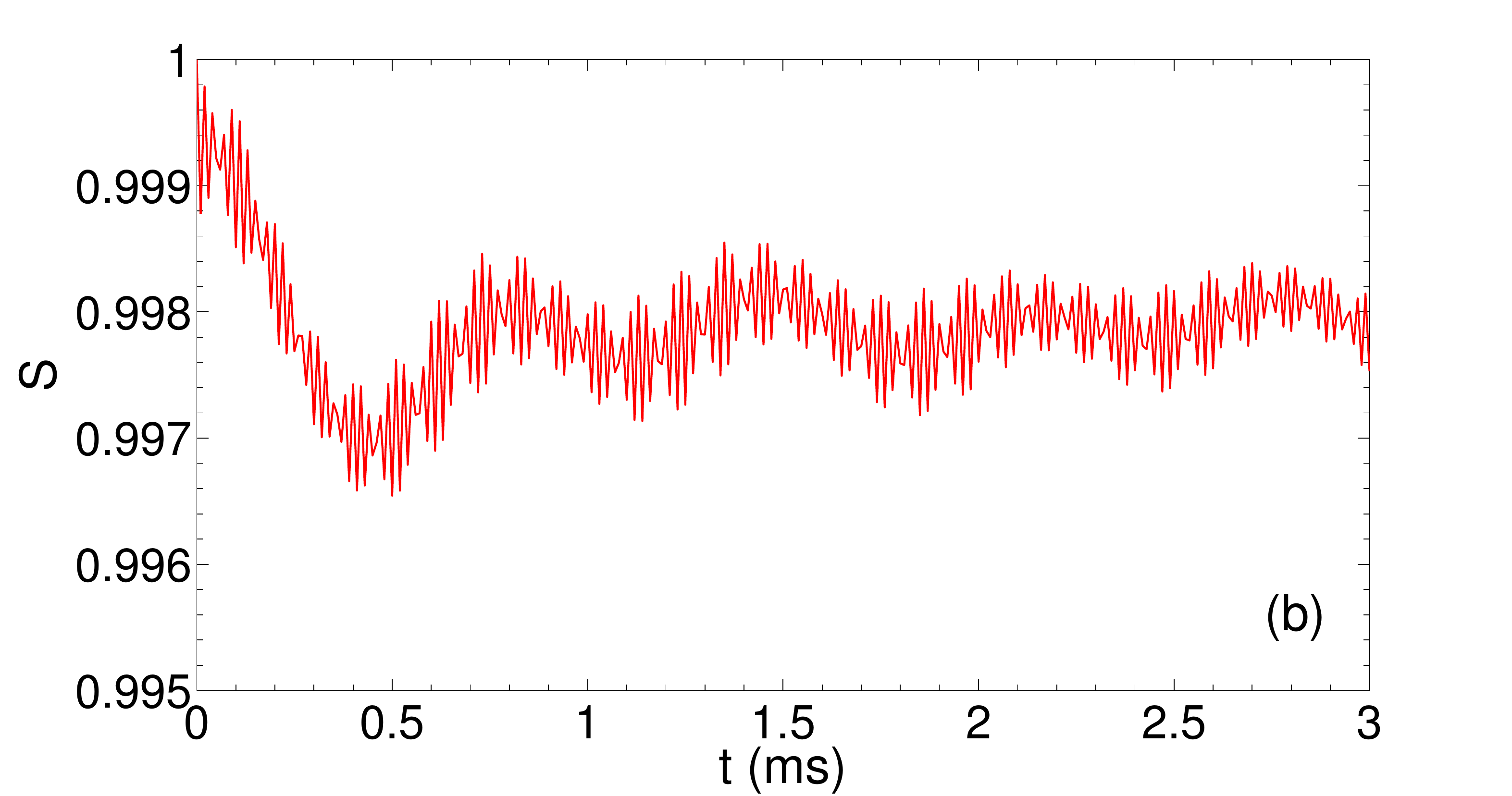}
\end{minipage}
\end{center}
\caption{(Color online) Decouple NV spin from the $\ele{13}{C}$ spin bath with continuous dynamical decoupling. The signal $S(t)=\bra{+}\rho(t)\ket{+}$ of NV spin as a function of time $t$. (a) In the example of $\ele{1}{H}_3 \ele{31}{P} \mbox{O}_4$, the magnetic field is $\gamma_{\ele{31}{P}}B=500\mbox{kHz}$ (i.e. $B=290\mbox{G}$, $\gamma_{\ele{13}{C}}B=310.6\mbox{kHz}$), and the driving amplitude is $\Omega=500\mbox{kHz}$. (b) In the example of $\ele{1}{H}_2\mbox{O}$, the magnetic field is $\gamma_{\ele{1}{H}}B=500\mbox{kHz}$ (i.e. $B=117\mbox{G}$, $\gamma_{\ele{13}{C}}B=125.7\mbox{kHz}$), and the driving amplitude is $\Omega=400\mbox{kHz}$.}\label{C13}
\end{figure}

The NV center spin is coupled with a $\ele{13} {C}$ spin bath, the interaction is described as follows
\begin{equation}
H_{NV-\ele{13}{C}}= \frac{\hbar\mu_0}{4\pi} \sum_{m} \frac{\gamma_m\gamma_e}{R_{m}^3} \blb{{\bf S} \cdot {\bf I}_m-3\bla{{\bf S}\cdot \hat{r}}\bla{{\bf I}_m\cdot \hat{r}}},
\end{equation}
where $R_{m}$ is the distance from the NV center to the nuclei, and $\hat{r}$ is the unit vector that connects the NV center and nuclei. The Hamiltonian of the spin bath itself is
\begin{equation}
H_{\ele{13}{C}}= -\sum_n\gamma_n {\bf B}\cdot {\bf I}_n +\frac{\hbar\mu_0}{4\pi} \sum_{n>m} \frac{\gamma_m\gamma_n}{R_{mn}^3} \blb{{\bf I}_m\cdot {\bf I}_n-3\bla{{\bf I}_m\cdot \hat{r}}\bla{{\bf I}_n\cdot \hat{r}}}.
\end{equation}
where $R_{mn}$ is the distance between two nuclei, and $\hat{r}$ is the unit vector that connects two nuclei. As we want to single out the effect of the target system, we apply continuous driving field on the NV spin which satisfies the Hatmann-Hahn condition for the target nuclear spin. Here, we use exact numerical simulation to show that continuous dynamical decoupling is very efficient for the parameters that we use for the examples in the main text, see Fig.\ref{C13}. Due to the computational overhead, we only consider a diamond sample with $8$ $\ele{13}{C}$ spins in a 4 nm sphere to demonstrate the essential idea.  We remark that, in a similar way, continuous dynamical decoupling can also suppress the noise from the surface of diamond if it is due to the spins that are different from the target spin.

{\it Acknowledgements} The work was supported by the Alexander von Humboldt Foundation, the EU Integrating Project Q-ESSENCE, the EU STREP PICC and DIAMANT, the BMBF Verbundprojekt QuOReP, DFG (FOR 1482, FOR 1493, SFB/TR 21) and DARPA. J.-M.C was supported also by a Marie-Curie Intra-European Fellowship within the 7th European Community Framework Programme.  We acknowledge the bwGRiD for computational resources.

\emph{Note:} After we finished this manuscript, three relevant experiments were published with the demonstration of sensing distant carbon-13 nuclear spins inside diamond \cite{Kolk1204,Zhao1204,Tam1205}.


\begin{thebibliography}{99}

\bibitem{Orrit99} W. E. Moerner and Michel Orrit, \emph{Illuminating Single Molecules in Condensed Matter}, Science {\bf 283}, 1670 (1999).

\bibitem{Silbey04} E. Barkai, Y. Jung and R. Silbey, \emph{Theory of single-molecule spectroscopy: Beyond the Ensemble Average}, Ann. Rev. of Phys. Chem. {\bf 55}, 457 (2004).

\bibitem{Lemke08} A. A. Deniz, S. Mukhopadhyay and E. A. Lemke, \emph{Single-molecule biophysics: at the interface of biology, physics and chemistry}, J. R. Soc. Interface {\bf 5}, 15 (2008).

\bibitem{Lord10} S. J. Lord, H. D. Lee and W. E. Moerner, \emph{Single-Molecule Spectroscopy and Imaging of Biomolecules in Living Cells}, Anal. Chem. {\bf 82} (6), 2192 (2010).

\bibitem{Twa02} W. Harneit, C. Meyer, A. Weidinger, D. Suter, J. Twamley, \emph{Architectures for a Spin Quantum Computer Based on Endohedral Fullerenes}, Phys. Stat. Sol. {\bf 3}, 453 (2002).

\bibitem{Ben06} S. C Benjamin, A. Ardavan, G. A. D. Briggs, D. A. Britz, D. Gunlycke, J. Jefferson, M. A. G. Jones, D. F. Leigh, B. W. Lovett, A. N. Khlobystov, S. A. Lyon, J. J. L. Morton, K. Porfyrakis, M. R. Sambrook and A. M. Tyryshkin, \emph{Towards a fullerene-based quantum computer}, J. Phys. Condens. Matter {\bf 18}, S867 (2006).

\bibitem{Kane98} B. E. Kane, \emph{A silicon-based nuclear spin quantum computer}, Nature {\bf 393}, 133 (1998).

\bibitem{Maze08} J. R. Maze, P. L. Stanwix, J. S. Hodges, S. Hong, J. M. Taylor, P. Cappellaro, L. Jiang, M. V. Gurudev Dutt, E. Togan, A. S. Zibrov, A. Yacoby, R. L. Walsworth, M. D. Lukin, \emph{Nanoscale magnetic sensing with an individual electronic spin in diamond}, Nature {\bf 455}, 644 (2008).

\bibitem{Bal08} G. Balasubramanian, I. Y. Chan, R. Kolesov, M. Al-Hmoud, J. Tisler, C. Shin, C. Kim, A. Wojcik, P. R. Hemmer, A. Krueger, T.Hanke, A. Leitenstorfer, R. Bratschitsch, F. Jelezko, J. Wrachtrup, \emph{Nanoscale imaging magnetometry with diamond spins under ambient conditions}, Nature {\bf 455}, 648 (2008).

\bibitem{Hall09} L. T. Hall, J. H. Cole, C. D. Hill, L. C. L. Hollenberg, \emph{Sensing of Fluctuating Nanoscale Magnetic Fields Using Nitrogen-Vacancy Centers in Diamond}, Phys. Rev. Lett. {\bf 103}, 220802 (2009).

\bibitem{Hanson11} G. de Lange, D. Riste, V. V. Dobrovitski, and R. Hanson, \emph{Single-Spin Magnetometry with Multipulse Sensing Sequences}, Phys. Rev. Lett. {\bf 106}, 080802 (2011).

\bibitem{Sch11} M. Schaffry, E. M. Gauger, J. J. L. Morton, and S. C. Benjamin, \emph{Proposed Spin Amplification for Magnetic Sensors Employing Crystal Defects}, Phys. Rev. Lett. {\bf 107}, 207210 (2011).

\bibitem{Hall10} L. T. Hall, C. D. Hill, J. H. Cole, B. St\"{a}dler, F. Caruso, P. Mulvaney, J. Wrachtrup, and L. C. L. Hollenberg, \emph{Monitoring ion-channel function in real time through quantum decoherence},Proc. Natl. Acad. Sci. U.S.A. {\bf 107}, 18777 (2010).

\bibitem{Cole09} J. H. Cole, L. C. L. Hollenberg, \emph{Scanning quantum decoherence microscopy}, Nanotechnology {\bf 20} 495401 (2009).

\bibitem{Yacoby11} M. S. Grinolds, P. Maletinsky, S. Hong, M. D. Lukin, R. L Walsworth, A. Yacoby, \emph{Quantum control of proximal spins using nanoscale magnetic resonance imaging}, Nature Physics {\bf 7}, 687 (2011).

\bibitem{Fu07} C.-C. Fu, H.-Y. Lee, K. Chen, T.-S. Lim, H.-Y. Wu, P.-K. Lin, P.-K. Wei, P.-H. Tsao, H.-C. Chang, \emph{Characterization and application of single fluorescent nanodiamonds as cellular biomarkers}, Proc. Natl. Acad. Sci. U.S.A. {\bf 104}, 727 (2007).

\bibitem{Chao07} J.-I. Chao, E. Perevedentseva, P.-H. Chung, K.-K. Liu, C.-Y. Cheng, C.-C. Chang, and C.-L. Cheng, \emph{Nanometer-Sized Diamond Particle as a Probe for Biolabeling}, Biophys. J. {\bf 93}, 2199 (2007).

\bibitem{Mcg11} L. P. McGuinness,	Y. Yan, A. Stacey,	 D. A. Simpson, L. T. Hall, D. Maclaurin, S. Prawer, P. Mulvaney, J. Wrachtrup, F. Caruso, R. E. Scholten, and L. C. L. Hollenberg, \emph{Quantum measurement and orientation tracking of fluorescent nanodiamonds inside living cells}, Nature Nanotechnology {\bf 6}, 358 (2011). 

\bibitem{Grotz11} B. Grotz, J. Beck, P. Neumann, B. Naydenov, R. Reuter, F. Reinhard, F. Jelezko, J. Wrachtrup, D. Schweinfurth, B. Sarkar and P. Hemmer, \emph{Sensing external spins with nitrogen-vacancy diamond}, New J. Phys. {\bf 13}, 055004 (2011).

\bibitem{Childress06} L. Childress, M. V. Gurudev Dutt, J. M. Taylor, A. S. Zibrov, F. Jelezko, J. Wrachtrup, P. R. Hemmer and M. D. Lukin, \emph{Coherent Dynamics of Coupled Electron and Nuclear Spin Qubits in Diamond}, Science  {\bf 314}, 281 (2006).

\bibitem{Jiang09} L. Jiang, J. S. Hodges, J. R. Maze, P. Maurer, J. M. Taylor, D. G. Cory, P. R. Hemmer, R. L. Walsworth, A. Yacoby, A. S. Zibrov, M. D. Lukin, \emph{Repetitive Readout of a Single Electronic Spin via Quantum Logic with Nuclear Spin Ancillae}, Science  {\bf 326}, 267 (2009).

\bibitem{Neu10} P. Neumann, J. Beck, M. Steiner, F. Rempp, H. Fedder, P. R. Hemmer, J. Wrachtrup and F. Jelezko, \emph{Single-Shot Readout of a Single Nuclear Spin}, Science {\bf 329}, 542 (2010).

\bibitem{Liu10} N. Zhao, J.-L. Hu, S.-W. Ho, J. T. K. Wan, R. B. Liu, \emph{Atomic-scale magnetometry of distant nuclear spin clusters via nitrogen-vacancy spin in diamond}, Nature Nanotechnology {\bf 6}, 242 (2011).

\bibitem{Sar08} L. Cywinski, R. M. Lutchyn, C. P. Nave  and S. Das Sarma, \emph{How to enhance dephasing time in superconducting qubits}, Phys. Rev. B {\bf 77}, 174509 (2008).

\bibitem{Byl11} J. Bylander, S. Gustavsson, F. Yan, F. Yoshihara, K. Harrabi,	G. Fitch, D. G. Cory, Y. Nakamura, J.-S. Tsai, W. D. Oliver, \emph{Noise spectroscopy through dynamical decoupling with a superconducting flux qubit}, Nature Phys. {\bf 7}, 565 (2011).

\bibitem{Yuge11} T. Yuge, S. Sasaki, and Y. Hirayama, \emph{Measurement of the Noise Spectrum Using a Multiple-Pulse Sequence}, Phys. Rev. Lett. {\bf 107}, 170504 (2011).

\bibitem{Suter11b} G. A. \'{A}lvarez and D. Suter, \emph{Measuring the Spectrum of Colored Noise by Dynamical Decoupling}, Phys. Rev. Lett. {\bf 107}, 230501(2011).

\bibitem{Gill12} N. Bar-Gill, L. M. Pham, C. Belthangady, D. Le Sage, P. Cappellaro, J. R. Maze, M. D. Lukin, A. Yacoby and R. Walsworth, \emph{Suppression of spin bath dynamics for improved coherence of multi-spin-qubit systems}, Nat. Commun. {\bf 3}, 858 (2012).

\bibitem{Fone04} K.M. Fonesca-Romero, S. Kohler, P. H\"{a}nggi, \emph{Coherence control for qubits }, Chem. Phys. {\bf 296}, 307 (2004).

\bibitem{Pas04pra} P. Facchi, D. A. Lidar, and S. Pascazio, \emph{Unification of dynamical decoupling and the quantum Zeno effect}, Phys. Rev. A {\bf 69}, 032314 (2004).

\bibitem{Fan07pra} F. F. Fanchini, J. E. M. Hornos, and R. d. J. Napolitano, Phys. Rev. A {\bf 75}, 022329 (2007)

\bibitem{Tim08} N. Timoney, V. Elman, S. Glaser, C. Weiss, M. Johanning, W. Neuhauser, and Chr. Wunderlich, \emph{Error-resistant single-qubit gates with trapped ions}, Phys. Rev. A {\bf 77}, 052334 (2008).

\bibitem{Ber11} A. Bermudez, F. Jelezko, M. B. Plenio, A. Retzker, \emph{Electron-Mediated Nuclear-Spin Interactions between Distant Nitrogen-Vacancy Centers}, Phys. Rev. Lett. {\bf 107}, 150503 (2011).

\bibitem{Cai11} J.-M. Cai, B. Naydenov, R. Pfeiffer, L. P. McGuinness, K. D. Jahnke, F. Jelezko, M. B. Plenio, A. Retzker, \emph{Robust dynamical decoupling with concatenated continuous driving}, New J. Phys. {\bf 14}, 113023 (2012).

\bibitem{TimNature} N. Timoney, I. Baumgart, M. Johanning, A. F. Var\'{o}n, M. B. Plenio, A. Retzker
and Ch. Wunderlich, \emph{Quantum gates and memory using microwave-dressed states}, Nature {\bf 476}, 185 (2011).

\bibitem{Ste89} U.~E. Steiner and T. Ulrich, \emph{Magnetic field effects in chemical kinetics and related phenomena}, Chem. Rev \textbf{89}, 51 (1989).

\bibitem{Noji97} H. Noji, R. Yasuda, M. Yoshida, K. Kinosita Jr, \emph{Direct observation of the rotation of $\mbox{F}_1$-ATPase}, Nature {\bf 386}, 299 (1997).

\bibitem{Nish04} T. Nishizaka, K. Oiwa, H. Noji, S. Kimura, E. Muneyuki, M. Yoshida, K. Kinosita Jr, \emph{Chemomechanical coupling in F1-ATPase revealed by simultaneous observation of nucleotide kinetics and rotation}, Nature Structural and Molecular Biology {\bf 11}, 142 (2004).

\bibitem{ZXW07} M. D. Stone, M. Mihalusova, C. M. O'Connor, R. Prathapam, K. Collins, X.-W. Zhuang, 
\emph{Stepwise protein-mediated RNA folding directs assembly of telomerase ribonucleoprotein}, Nature {\bf 446}, 458 (2007).

\bibitem{Hahn62} S. R. Hartmann and E. L. Hahn, \emph{Nuclear Double Resonance in the Rotating Frame}, Phys. Rev. {\bf 128}, 2042 (1962). 

\bibitem{Taka08} S. Takahashi, R. Hanson, J. van Tol, M. S. Sherwin, D. D. Awschalom, \emph{Quenching Spin Decoherence in Diamond through Spin Bath Polarization}, Phys. Rev. Lett. {\bf 101}, 047601 (2008).

\bibitem{Okai12} B. K. Ofori-Okai, S. Pezzagna, K. Chang, R. Schirhagl, Y. Tao, B. A. Moores, K. Groot-Berning, J. Meijer, C. L. Degen, \emph{Spin Properties of Very Shallow Nitrogen Vacancy Defects in Diamond}, Phys. Rev. B {\bf 86}, 081406(R) (2012).

\bibitem{Ohno1207} K. Ohno, F. J. Heremans, L. C. Bassett, B. A. Myers, D. M. Toyli, A. C. B. Jayich, C. J. Palmstrom, D. D. Awschalom, \emph{Engineering shallow spins in diamond with nitrogen delta-doping},  	Appl. Phys. Lett. {\bf 101}, 082413 (2012).

\bibitem{Bala09} G. Balasubramanian, P. Neumann, D. Twitchen, M. Markham, R. Kolesov, N. Mizuochi, J. Isoya, J. Achard, J. Beck, J. Tissler, V. Jacques, P. R. Hemmer, F. Jelezko, J. Wrachtrup, \emph{Ultralong spin coherence time in isotopically engineered diamond}, Nature Materials {\bf 8}, 383 (2009).

\bibitem{Pezz10} S. Pezzagna, B. Naydenov, F. Jelezko, J. Wrachtrup, J. Meijer, \emph{Creation efficiency of nitrogen-vacancy centres in diamond}, New J. Phys. {\bf 12}, 065017 (2010).

\bibitem{SPREFMT} Peter G. Fajer in {\it Encyclopedia of Analytical Chemistry}, R.A. Meyers (Ed.) pp. 5725-5761, John Wiley and Sons Ltd, Chichester (2000).

\bibitem{Ritz00} T. Ritz, S. Adem, and K. Schulten, \emph{A Model for Photoreceptor-Based Magnetoreception in Birds}, Biophys. J. {\bf 78}, 707 (2000). 

\bibitem{Cai10prl} J.-M. Cai, G. G. Guerreschi, and H. J. Briegel, \emph{Quantum Control and Entanglement in a Chemical Compass}, Phys. Rev. Lett. {\bf 104}, 220502 (2010). 

\bibitem{Gau11prl} E. M. Gauger, E. Rieper, J. J. L. Morton, S. C. Benjamin, and V. Vedral, \emph{Sustained Quantum Coherence and Entanglement in the Avian Compass}, Phys. Rev. Lett. {\bf 106}, 040503 (2011).

\bibitem{Ritz11} T. Ritz, \emph{Quantum effects in biology: Bird navigation}, Procedia Chemistry 3, 262 (2011) 

\bibitem{Cai11prl} J.-M. Cai, \emph{Quantum Probe and Design for a Chemical Compass with Magnetic Nanostructures}, Phys. Rev. Lett. {\bf 106}, 100501 (2011).

\bibitem{Cai12pra} J.-M. Cai, F. Caruso, M. B. Plenio, \emph{Quantum limits for the magnetic sensitivity of a chemical compass}, Phys. Rev. A {\bf 85}, 040304(R) (2012).

\bibitem{Mour12} H. Mouritsen, P. J. Hore, \emph{The magnetic retina: light-dependent and trigeminal magnetoreception in migratory birds}, Current Opinion in Neurobiology {\bf 22}, 343 (2012).

\bibitem{HAB} R. Haberkorn, \emph{Density matrix description of spin-selective radical pair reactions}, Mol. Phys. 32, 1491 (1976).

\bibitem{Kolk1204} S. Kolkowitz, Q. P. Unterreithmeier, S. D. Bennett, M. D. Lukin, \emph{Sensing distant nuclear spins with a single electron spin}, Phys. Rev. Lett. {\bf 109}, 137601 (2012)

\bibitem{Zhao1204} N. Zhao, J. Honert, B. Schmid, J. Isoya, M. Markham, D. Twitchen, F. Jelezko, R.-B. Liu, H. Fedder, J. Wrachtrup, \emph{Sensing single remote nuclear spins}, Nature Nanotechnology {\bf 7}, 657–662 (2012).

\bibitem{Tam1205} T. H. Taminiau, J. J. T. Wagenaar, T. van der Sar, F. Jelezko, V. V. Dobrovitski, R. Hanson, \emph{Detection and control of individual nuclear spins using a weakly coupled electron spin}, Phys. Rev. Lett. {\bf 109}, 137602 (2012).

\end{thebibliography}
\end{document}